\documentstyle [aps,pre,epsf] {revtex}

\newcommand{\dc} {\delta c_0}
\newcommand{\eg} {{\it e.g., }}
\newcommand{\eps} {\epsilon}
\newcommand{\half} {\frac{1}{2}}
\newcommand{\ie} {{\it i.e., }}
\newcommand{\rmc} {{\rm c}}
\newcommand{\rmd} {{\rm d}}
\newcommand{\rme} {{\rm e}}
\newcommand{\rmr} {{\rm r}}
\newcommand{\vecn} {{\bf n}}
\newcommand{\vecr} {{\bf r}}
\newcommand{\vecR} {{\bf R}}

\def\tr{\mathop{\mbox{tr}}}

\begin{document}

\draft

\title
{Topography and instability of monolayers near domain boundaries}

\author{H.\ Diamant,$^1$ T.\ A.\ Witten,$^1$
C.\ Ege,$^2$ A.\ Gopal,$^2$ and K.\ Y.\ C.\ Lee$^2$ }
\address{$^1$James Franck Institute and Department of Physics,
University of Chicago,\\ 5640 South Ellis Avenue, Chicago, IL 60637 \\
$^2$Department of Chemistry and Institute for Biophysical
Dynamics, University of Chicago,\\ 5735 South Ellis Avenue, Chicago,
IL 60637}

\date{February 9, 2001}

\maketitle

\begin{abstract}
We theoretically study the topography of a biphasic surfactant
monolayer in the vicinity of domain boundaries. The differing
elastic properties of the two phases generally lead to a nonflat
topography of ``mesas", where domains of one phase are elevated
with respect to the other phase. The mesas are steep but low,
having heights of up to 10 nm. As the monolayer is laterally
compressed, the mesas develop overhangs and eventually become
unstable at a surface tension of about $K(\dc)^2$ ($\dc$ being the
difference in spontaneous curvature and $K$ a bending modulus). In
addition, the boundary is found to undergo a topography-induced
rippling instability upon compression, if its line tension is
smaller than about $K\dc$. The effect of diffuse boundaries on
these features and the topographic behavior near a critical point
are also examined. We discuss the relevance of our findings to
several experimental observations related to surfactant
monolayers: (i) small topographic features recently found near
domain boundaries; (ii) folding behavior observed in mixed
phospholipid monolayers and model lung surfactants; (iii)
roughening of domain boundaries seen under lateral compression;
(iv) the absence of biphasic structures
in tensionless surfactant films.
\end{abstract}



\section{Introduction}
\label{sec_intro}

Monolayers of amphiphilic molecules (surfactants) at water/air or
water/oil interfaces are used in numerous applications to reduce
interfacial tension, control wetting properties, stabilize
emulsions and foams, {\it etc.} \cite{Gelbart,Safran}. Monolayers
of biological surfactants (phospholipids) are commonly studied as
models for the surfaces of cell membranes and are also encountered
in various biological systems \cite{Birdi}. An important example
is the lung surfactant monolayer covering the alveoli in lungs,
whose main function is to lower the surface tension of the lungs,
thereby drastically reducing the mechanical work required for
breathing \cite{lung}.

Amphiphilic monolayers generally have a finite spontaneous
curvature arising from the asymmetry of the molecules as well as
the asymmetry with respect to electrostatic interactions (\ie the
differing dielectric properties of the polar and nonpolar phases
forming the interface) \cite{Glen}. Despite this tendency to bend,
homogeneous monolayers are almost always flattened by the
water/air or water/oil interfacial tension. Only for very low
(sometimes even negative \cite{MJP}) tension does a nonflat
conformation become energetically favorable for a homogeneous
monolayer. (This is achieved, \eg by extensive lateral
compression.) Such a reversible departure from a flat,
two-dimensional state to a three-dimensional conformation is
referred to as the {\em buckling transition} and has drawn
considerable attention \cite{MJP,Jalmes,Granek}. However, it is
not commonly observed in practice \cite{Jalmes}, since it is
usually preceded by other modes of collapse such as monolayer
breakage into multilayers \cite{multilayer1,multilayer2} 
and ejection of vesicles or micelles \cite{ejection}.
The possibility to explore the third dimension upon compression is
of particular interest in the case of lung surfactant monolayers,
which are required to change their projected area significantly
during the compression-expansion cycle of breathing.

The two-dimensional fluid comprising a monolayer may separate into
domains of different coexisting phases. Single-component
monolayers exhibit gas/liquid-expanded,
liquid-expanded/liquid-condensed and liquid-condensed/solid
coexistence \cite{Gelbart,Safran,Birdi}, whereas mixed monolayers
may form domains of differing composition. A special, well-studied
property of surfactant monolayers is the stabilization of finite
domains and modulated phases due to long-range electrostatic
interactions \cite{modulated1,modulated2}. The coupling between
lateral variations in composition and curvature was thoroughly
studied as well
\cite{Granek,Leibler,Japan,Wang,Fred93,Komura,Harden,German},
mainly with regard to various domain structures on surfaces and
shape transformations of bilayer vesicles.

\begin{figure}
\centerline{\epsfxsize=0.8\textwidth
\hbox{\epsffile{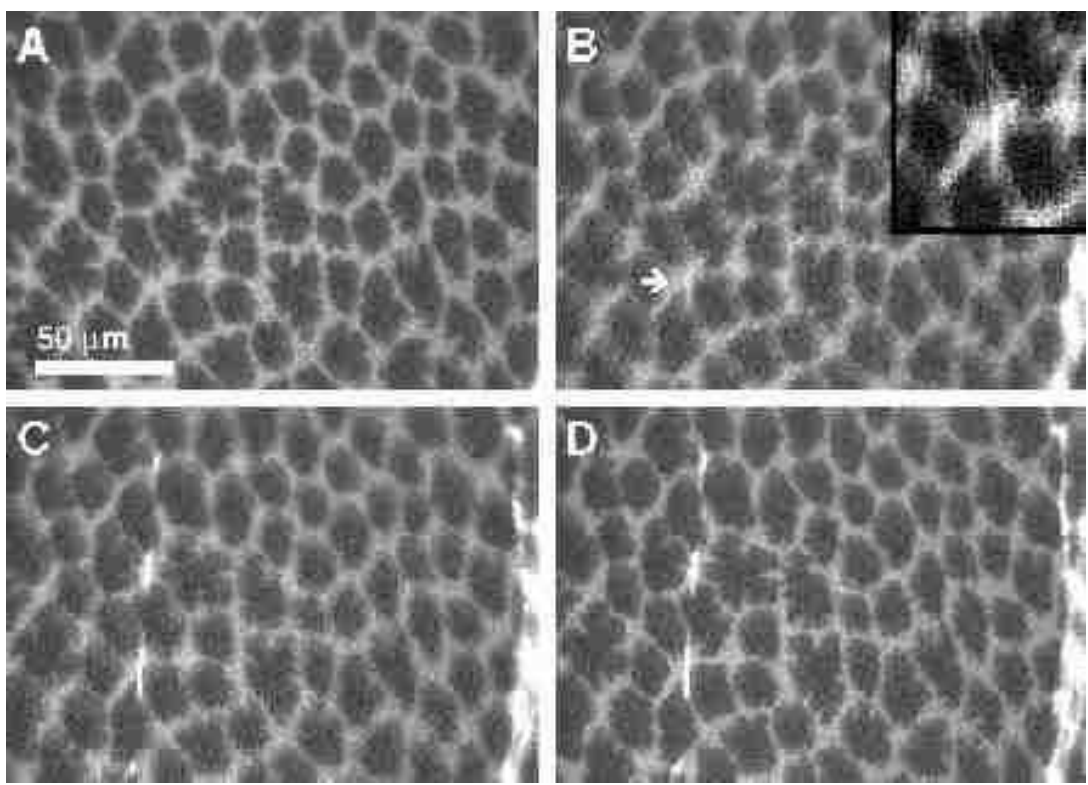}}}
\vspace{.5cm}
\caption[]{Fluorescence microscopy images of the
folding instability. A) Section of the monolayer just before
folding ($t=0$), exhibiting the biphasic domain structure. Dark
regions are DPPC-rich; bright ones are POPG-rich. B) The same
section at $t=$1/30 s. A micron-scale fold appears in between
domain walls (indicated by arrow). The image is blurred because of
monolayer movement during folding. The inset shows a
contrast-enhanced image of the fold, magnified by 50 percent. C)
The fold at $t=$2/30 s, having propagated to nearby domains. D)
The fold at $t=$4/30 s, after the fast monolayer movement has
ceased.} 
\label{fig_fold}
\end{figure}

\begin{figure}
\centerline{\epsfxsize=0.8\textwidth
\hbox{\epsffile{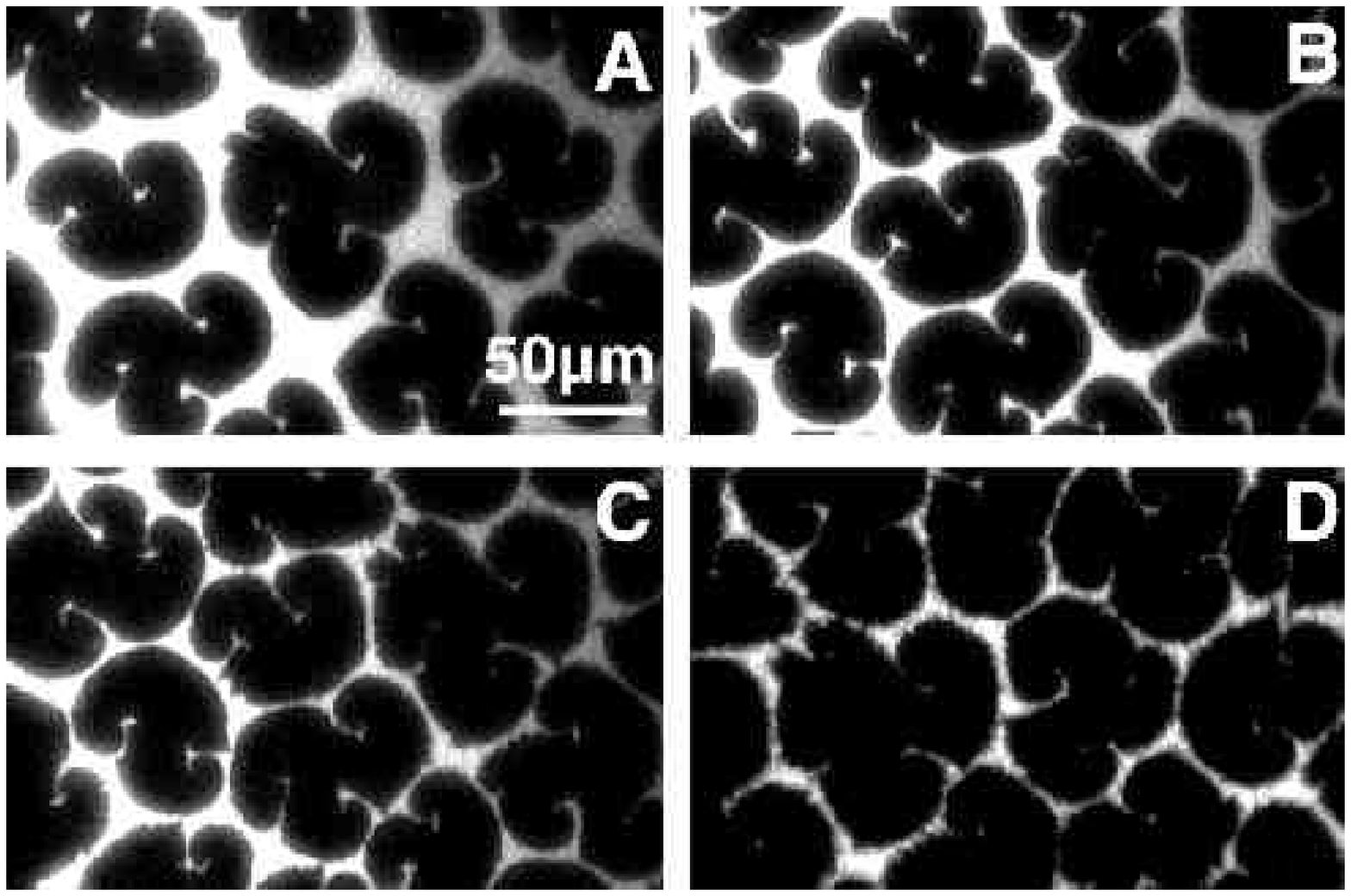}}}
\vspace{.5cm}
\caption[]{Roughening of domain boundaries upon compression,
as observed by fluorescence microscopy.
The monolayer consists of DPPC phospholipids and
lies over an aqueous solution of 0.2M NaCl at temperature
24.5$^\circ$C.
Dark regions correspond to a liquid-condensed phase and
bright ones to a liquid-expanded phase.
The lateral pressures and molecular areas (in mN/m
and \AA$^2$, respectively) are:
A) 16.9, 54.8;
B) 19.4, 53.0;
C) 21.6, 51.9;
D) 23.4, 51.2.
Images were contrast-enhanced.
}
\label{fig_ripple_exp}
\end{figure}

Despite extensive research on surfactant monolayers there are
important features, in particular of biphasic monolayers, which
are not well understood. Recent experiments on mixed phospholipid
monolayers have revealed a new type of local folding upon
compression, which is believed to be important for the function of
lungs \cite{KaYee} (see Fig.~\ref{fig_fold}).
Another observed feature is the appearance of
rough domain boundaries upon compression \cite{Canay}
(see Fig.~\ref{fig_ripple_exp}).

In the current work we study the relation between lateral domain
structure and monolayer topography in more detail, focusing on the
conformation of monolayers in the vicinity of domain boundaries
\cite{ourEPL}. We thereby try to shed some light on the
unexplained features mentioned above. Domains of different density
and/or composition in a biphasic monolayer should generally have
differing elastic properties. The requirements of mechanical
equilibrium and smoothness of the monolayer surface lead to
nonflat conformations attempting to ``reconcile" the different
properties of the contiguous domains. These simple mechanical
considerations result in a surprisingly rich behavior, including
the formation of overhangs and emergence of instabilities, as
discussed in the following sections.

The Monge representation and linearization of profile equations
have been ubiquitously used in the theoretical modeling of
monolayers and membranes \cite{Safran}. These mathematical
simplifications describe the topography by a single-valued height
function assumed to have moderate slopes. By contrast, the
phenomena discussed in the current work involve, in an essential
way, steep slopes and overhangs. We thus avoid the Monge
representation and solve the nonlinear profile equations. In order
for the mathematics to remain tractable we resort to another
simplification---the profile is assumed to be uniform along one
lateral direction (namely, the direction parallel to the domain
boundary)---rendering the nonlinear equations one-dimensional.
This constraint is further discussed in the next section; it is
partially relaxed in the treatment of boundary rippling in
Sec.~\ref{sec_rippling}.

The basic model and its assumptions are presented in
Sec.~\ref{sec_model}. We then proceed in Sec.~\ref{sec_sharp} to
review the simplest case of an infinitely sharp, straight domain
boundary \cite{ourEPL}. Despite its simplicity, this limiting case
demonstrates most of our qualitative results. The calculation is
refined in Sec.~\ref{sec_diffuse} where a boundary of nonzero
thickness is considered. In Sec.~\ref{sec_rippling} we study the
stability of a straight domain boundary to lateral undulations.
The topographic behavior of a monolayer near its critical point is
examined in Sec.~\ref{sec_Tc}. Finally, in
Sec.~\ref{sec_discussion}, we discuss the various results and
their relevance to experiments.


\section{The Model}
\label{sec_model}

\begin{figure}
\centerline{\epsfxsize=0.45\textwidth \hbox{\epsffile{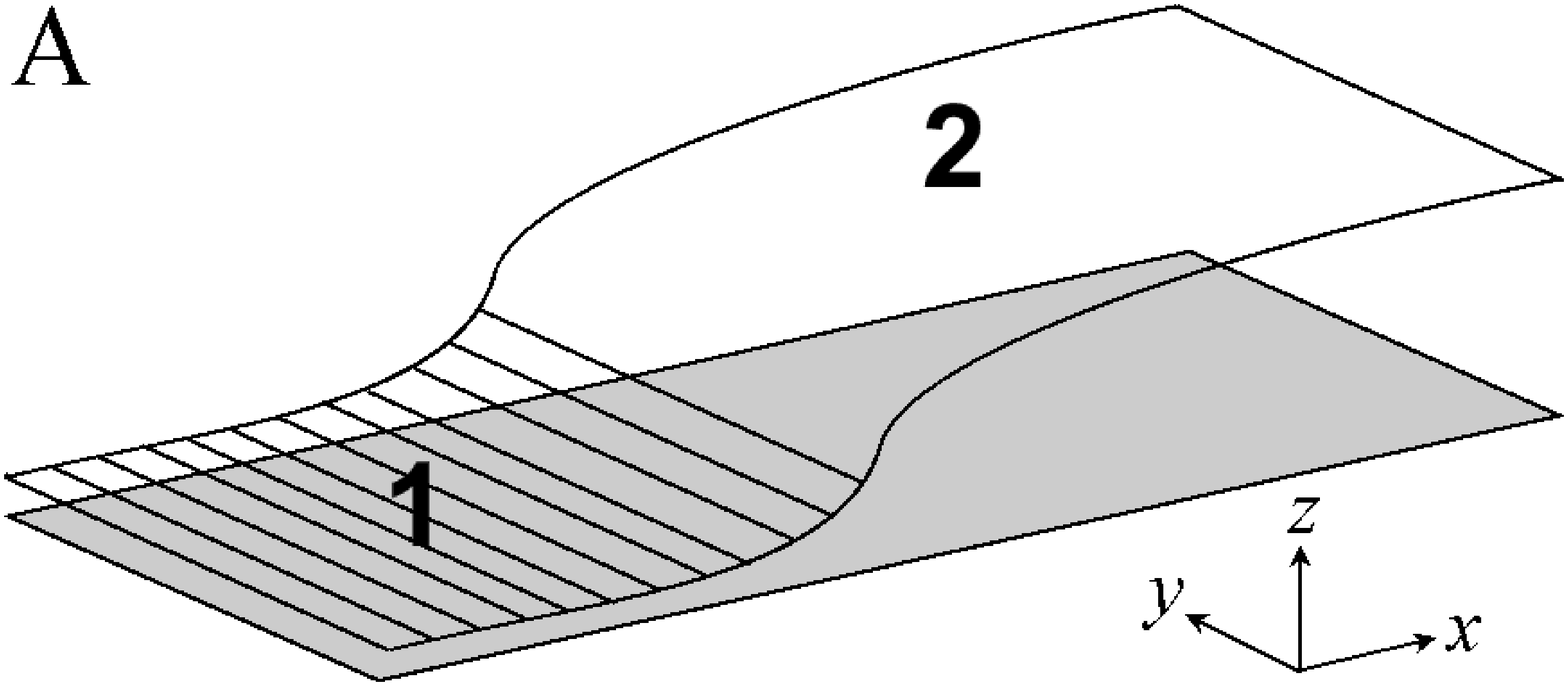}}
\hspace{.5cm} \epsfxsize=0.45\textwidth
\hbox{\epsffile{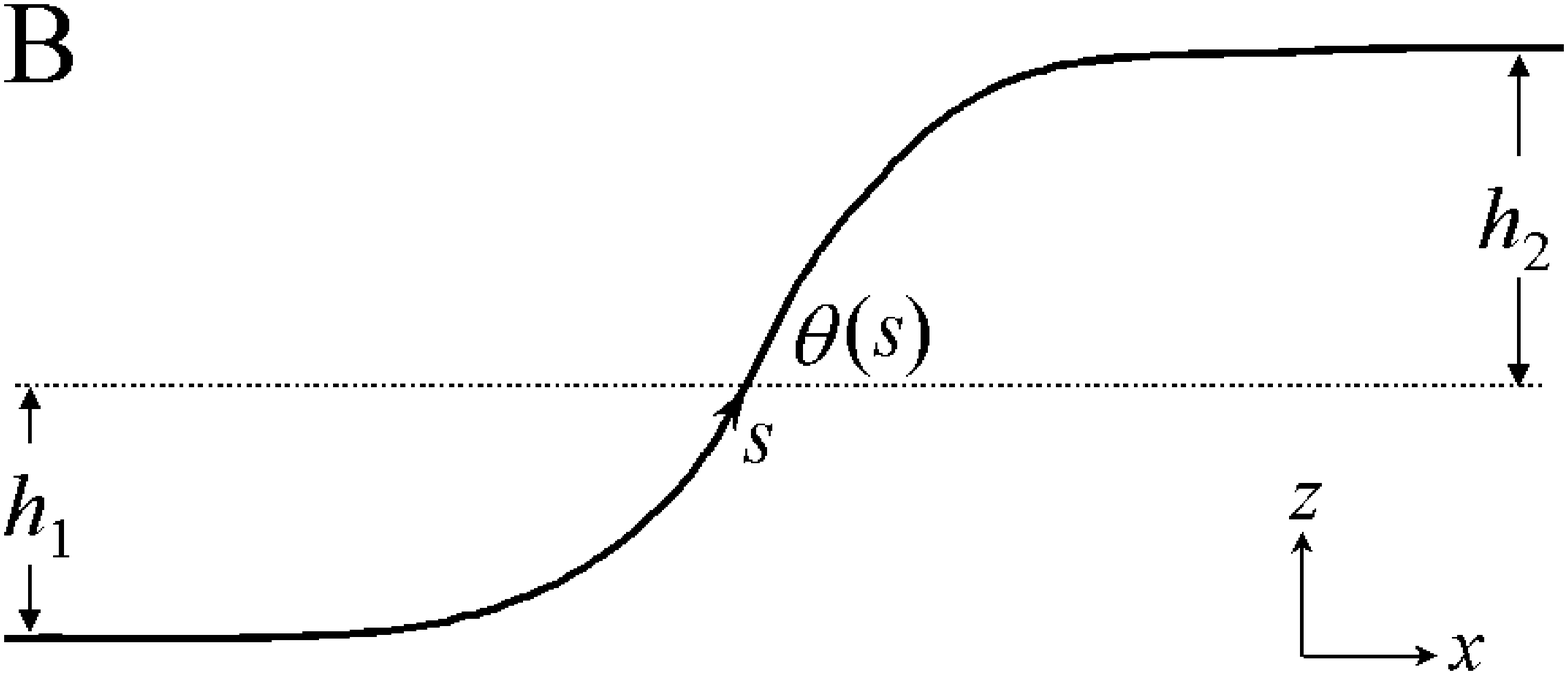}}}
\vspace{.5cm} \caption[]{A) Schematic sketch of the monolayer in
the boundary region. A boundary lying parallel to the $y$ axis
separates two large domains denoted by 1 and 2. B) Cross-section
parallel to the $xz$ plane. The monolayer conformation is
parameterized by the angle $\theta(s)$ it makes with the $xy$
reference plane at curvilinear distance $s$ from the boundary.}
\label{fig_scheme}
\end{figure}

Four length scales are distinguished when studying the elasticity
of a biphasic monolayer: the typical domain size, $L$, the width
of a domain boundary, $\xi$, the typical spontaneous radius of
curvature, $c_0^{-1}$, and the elastic length,
$\lambda=(K/\gamma)^{1/2}$, determining the lateral extent of
height variations ($K$ being the bending modulus and $\gamma$ the
surface tension). An important observation is that in most
practical cases the domain size is much larger than all other
length scales---$L$ is typically of order 10 $\mu$m, whereas
$\xi$, $c_0^{-1}$ and $\lambda$ are of order 1--10 nm. This allows
us to focus on a single, straight boundary between two large
domains and regard the centers of the domains as infinitely far
away. We thus represent the boundary region as a surface whose far
left and far right have different spontaneous curvatures, $c_{01}$
and $c_{02}$ \cite{ft_diffK}. (Throughout this paper we assume,
without loss of generality, $c_{01}>c_{02}$.) The surface is
uniform in the $y$ direction parallel to the boundary but can
curve in the perpendicular $x$ direction, as depicted in
Fig.~\ref{fig_scheme}A.

In fact, an inflected conformation perpendicular to the boundary
is a straightforward consequence of such a lateral structure. Far
away from the boundary (\ie at the centers of the two contiguous
domains) the surface is flat. Because of the
nonzero spontaneous curvature the far left side is subjected
to a bending moment of $Kc_{01}$, supplemented by a tensile torque
of $-\gamma h_1$, where $-h_1$ is the height of this side relative
to the boundary (see Fig.~\ref{fig_scheme}B). Similarly, a moment
of $Kc_{02}+\gamma h_2$ is exerted on the far right side.
Mechanical equilibrium requires that these two moments balance
each other, \ie
\begin{equation}
  h = h_1 + h_2 = {K\dc}/{\gamma} = \lambda^2\dc,
\label{sumrule}
\end{equation}
where $\dc\equiv c_{01}-c_{02}$. Thus, an inflected conformation
with a finite height difference occurs for any finite $\dc$ and
$\gamma$. A measure of the inflection sharpness is
$h/\lambda=\lambda\dc$. As compression increases (\ie $\gamma$
decreases \cite{ft_presstens}), the inflection becomes higher and
sharper. Note that the integrated height difference is insensitive
to details of the inner boundary region. Equation (\ref{sumrule})
can therefore serve as a rigorous sum rule for more detailed
calculations such as those presented in Secs.~\ref{sec_sharp} and
\ref{sec_diffuse}.

The elastic energy to be minimized in
order to find the monolayer topography \cite{ft_fluct} is
\begin{equation}
  G = \int_A\rmd A\left(\half Kc^2 - Kc_0c\right) +
  \gamma\int_A\rmd(A-A_{\rm p}) + \tau\int_\vecR\rmd (R-y),
\label{G_general}
\end{equation}
where $A$ denotes the monolayer surface, $A_{\rm p}$ its
projection onto the $xy$ reference plane, $c$ the local surface
curvature, $\vecR$ the trajectory of the domain boundary, and
$\tau$ the line tension of the boundary. The functional
(\ref{G_general}) has been defined such that a flat surface with a
straight domain boundary running parallel to the $y$ axis is the
reference state of zero energy. Since the surface we consider
in Secs.~\ref{sec_sharp} and \ref{sec_diffuse} is
uniform in the $y$ direction, we may represent it by the local
angle $\theta(s)$ it makes with the reference $xy$ plane at curvilinear
distance $s$ from the boundary (see Fig.~\ref{fig_scheme}B). The
energy functional (\ref{G_general}) is then rewritten as
\begin{equation}
  g[\theta(s)] \equiv \frac{G}{L} = \int_{-\infty}^\infty\rmd s
  \left[ \half K \dot{\theta}^2 - Kc_0\dot{\theta} +
  \gamma(1-\cos\theta) \right],
\label{g_general}
\end{equation}
where $L$ is the length of the boundary, a dot denotes $\rmd/\rmd
s$, and $c_0$ is not regarded as a constant but varies with $s$.
In order for the surface to be smooth everywhere we require that
$\theta(s)$ be a continuous function \cite{ft_kink}.
It is useful to notice that, despite the linear term in $\dot\theta$, 
the functional (\ref{g_general}) is invariant under $s$-reversal,
$s\rightarrow -s$.
This is because we can break any plausible choice of $c_0(s)$
into a constant, $(c_{01}+c_{02})/2$, plus an odd
function of $s$ and, assuming that the surface becomes flat far away
from the boundary, the constant term does not contribute to the
integral.
In other words, we may specialize to the case $c_{01}=-c_{02}$.
Hence, minimization will necessarily produce symmetric angle
profiles $\theta(s)$ and antisymmetric topographies $h(s)$.
(Note that this argument does not hold if $K$ is taken as
nonuniform \cite{ourEPL}.)


\section{Sharp domain boundary}
\label{sec_sharp}

We begin with the simple case where the boundary thickness $\xi$
is much smaller than the other length scales, extending our
earlier work \cite{ourEPL}.
(We show in Sec.~\ref{sec_diffuse} that the condition for this
sharp limit is in fact weaker, the requirement being merely
$\xi\dc\ll 1$.)
In this limit the boundary may be regarded as infinitely sharp,
accompanied by a step-function jump in spontaneous curvature,
\[
  c_0 = \left\{ \begin{array}{ll}
  c_{01} \ \ \ \ & s<0 \\
  c_{02} & s>0. \end{array}\right.
\]
Substituting this spatial variation of $c_0$ in
Eq.~(\ref{g_general}) we can integrate the linear term in
$\dot{\theta}$. The energy is then rewritten as
\begin{equation}
  g[\theta(s)] = \int_{-\infty}^\infty\rmd s \left[
  \half K \dot{\theta}^2 + \gamma(1-\cos\theta) \right]
  - K\dc\theta_0,
\label{g_sharp1}
\end{equation}
where $\theta_0\equiv\theta(s=0)$ is the maximum inflection angle.

The integral in Eq.~(\ref{g_sharp1}) has the familiar form of the
Sine-Gordon action. Variation with respect to $\theta(s\neq 0)$ gives
a Sine-Gordon profile equation,
\begin{equation}
  \ddot{\theta} = \lambda^{-2}\sin\theta.
\label{SG_sharp}
\end{equation}
First integration of Eq.~(\ref{SG_sharp}) yields
\begin{equation}
  \dot{\theta} = \left\{ \begin{array}{ll}
  2\lambda^{-1}\sin(\theta/2) \ \ \ \ & s<0 \\
  -2\lambda^{-1}\sin(\theta/2) & s>0.
  \end{array}\right.
\label{tdot_sharp}
\end{equation}
In the current, simple case, due to the boundary conditions
$\theta(s\rightarrow\pm\infty)=0$, second integration can be
carried out as well. The following soliton profile
is obtained:
\begin{equation}
  \tan\frac{\theta}{4} =
  \tan\frac{\theta_0}{4}\exp(-|s|/\lambda).
\label{prof_sharp}
\end{equation}
Finally, we need a condition for the jump in curvature at the
sharp boundary. This is derived mathematically by either
integrating $\ddot\theta$ of Eq.~(\ref{SG_sharp}) over an
infinitesimal length around the boundary, or taking the variation
of $g$ with respect to $\theta_0$. The same result is obtained,
nonetheless, by a simple moment-balance argument: the bending
moment acting on the boundary from the left,
$K[\dot{\theta}(0^-)-c_{01}]$, must balance the one acting from
the right, $K[\dot{\theta}(0^+)-c_{02}]$. Hence,
\[
  \dot{\theta}(0^-) - \dot{\theta}(0^+) = \dc,
\]
which determines $\theta_0$ as
\begin{equation}
  \sin\frac{\theta_0}{2} = \frac{\lambda\dc}{4} = \frac{h}{4\lambda}.
\label{t0_sharp}
\end{equation}
The dependence of the slope on the ratio $h/\lambda$ in the
sharp-boundary limit readily follows from dimensional analysis.
Yet, the exact nonlinear dependence given by Eq.~(\ref{t0_sharp})
is essential for our main results, as will be demonstrated below.

Thus, we infer again that for any finite $\dc$ and $\gamma$
the monolayer attains an inflected shape whose profile is
given by Eqs.~(\ref{prof_sharp}) and (\ref{t0_sharp}).
Integrating $\int_{-\infty}^\infty\rmd s\sin\theta(s)$, one
verifies that the general sum rule for the total height difference,
Eq.~(\ref{sumrule}), is satisfied.
The energy of the inflected conformation is calculated by
substituting the obtained profile back in Eq.~(\ref{g_sharp1}),
\begin{equation}
  g(\theta_0) = K\dc \left( 2\tan\frac{\theta_0}{4} - \theta_0 \right),
\label{g_sharp2}
\end{equation}
which, as expected, is negative for $\theta_0<\pi$, \ie the
inflected shape is favored over the flat one. We also calculate
the projected area as a function of tension, either by
integration, $A_{\rm p}=L\int_{-\infty}^\infty\rmd s\cos\theta$,
or by the following derivative,
\begin{equation}
  \Delta L \equiv \frac{A-A_{\rm p}}{L} =
  \frac{\partial g}{\partial\gamma} =
  \frac{32}{\dc} \sin\frac{\theta_0}{2} \sin^2\frac{\theta_0}{4},
\end{equation}
and the lateral compressibility,
\begin{equation}
  C = -\frac{\partial\Delta L}{\partial\gamma} =
  \frac{1}{K(\dc)^3} \frac{[4\sin(\theta_0/2)]^3 \sin(3\theta_0/2)}
  {\cos(\theta_0/2)}.
\end{equation}

\begin{figure}
\centerline{\epsfxsize=0.5\textwidth
\hbox{\epsffile{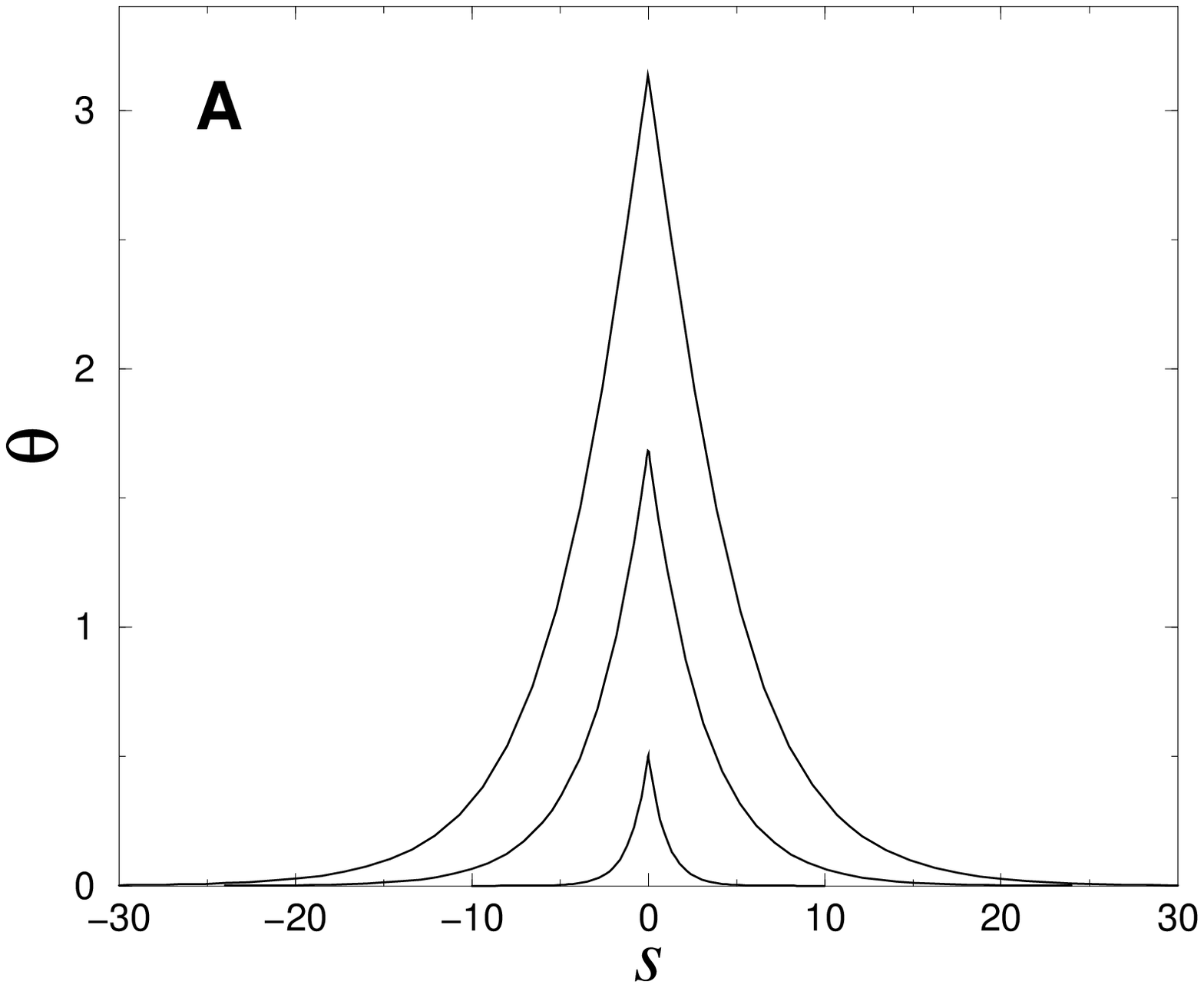}} \epsfxsize=0.5\textwidth
\hbox{\epsffile{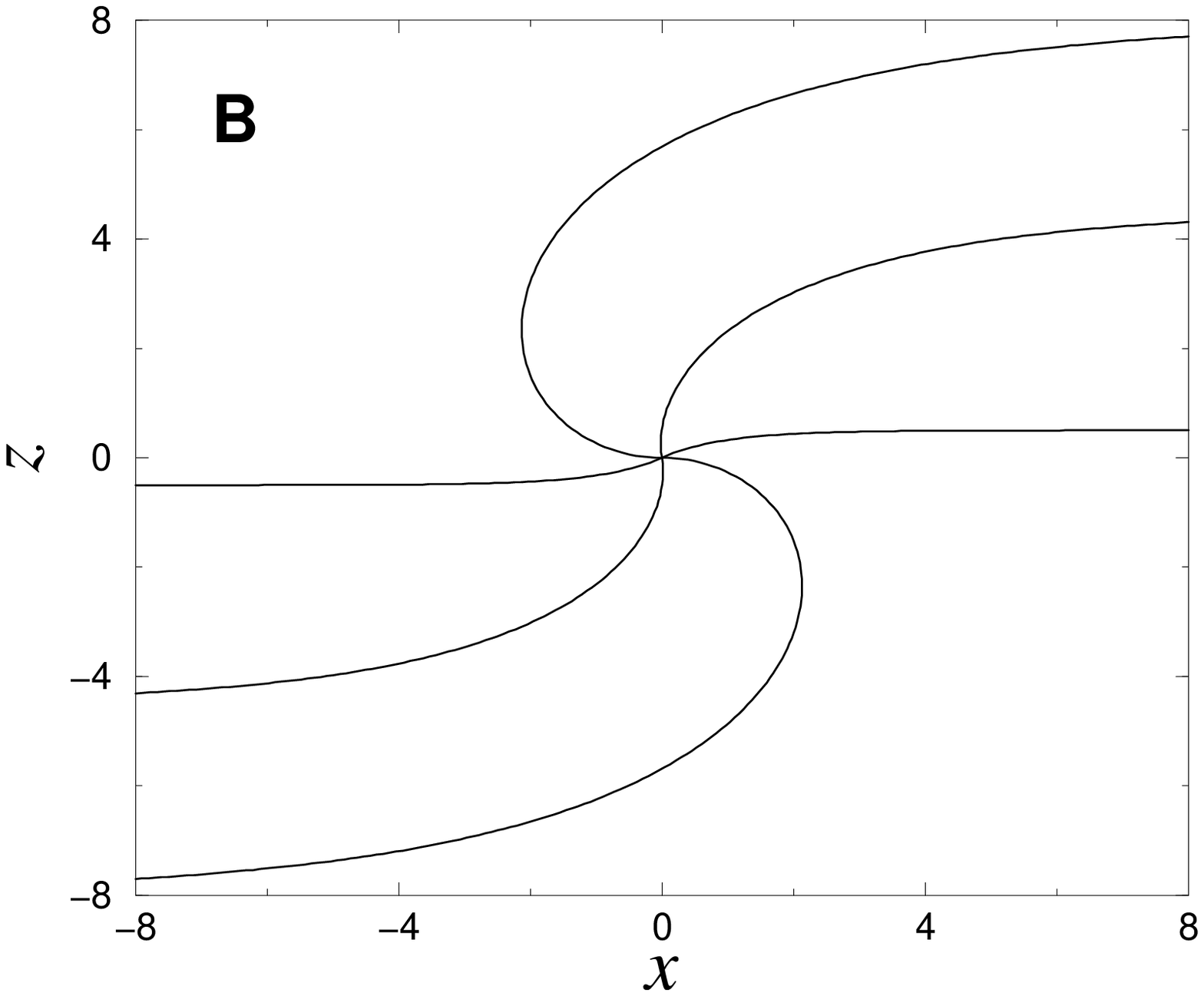}}}
\caption[]{A) Slope angle profiles near a sharp domain boundary as
compression is increased. The curves are obtained from Eqs.\
(\ref{prof_sharp}) and (\ref{t0_sharp}) using the values (from
bottom to top) $\lambda\dc=1, 3, 4$. B) The corresponding spatial
conformations. All lengths are given in units of $(\dc)^{-1}$.}
\label{fig_compress}
\end{figure}

As the monolayer is compressed by progressively decreasing $\gamma$
or increasing $\Delta L$ (depending on the experimental setup),
the inflected
profile becomes sharper (larger $\theta_0$, Eq.~(\ref{t0_sharp})),
higher (larger $h$, Eq.~(\ref{sumrule})) and more favorable
(lower $g$, Eq.~(\ref{g_sharp2})). The process is demonstrated
in Fig.~\ref{fig_compress}. For $\gamma<\frac{1}{8}K(\dc)^2$
the inflection angle $\theta_0$ becomes larger than $\pi/2$
and a stable overhang forms.
However, beyond a critical value of compression,
\begin{equation}
  \gamma < \gamma_\rmc = \frac{1}{16}K(\dc)^2, \ \ \ \
  \Delta L > \Delta L_\rmc = \frac{16}{\dc},
\label{critical_sharp}
\end{equation}
Eq.~(\ref{t0_sharp}) has no solution. This occurs in the current
calculation when $\theta_0=\theta_{0\rmc}=\pi$, at which point the
inflection remains finite, $h=h_\rmc=16/\dc$, yet the lateral
compressibility diverges,
\begin{equation}
  C\sim(\pi-\theta_0)^{-1}\sim(\gamma-\gamma_\rmc)^{-1/2}.
\end{equation}
Note that this instability is revealed only in the nonlinear
representation of the elastic problem (\eg {\it via}
Eq.~(\ref{t0_sharp}) which is nonlinear in $\theta_0$). A theory
relying on the Monge representation and moderate slopes would
inevitably miss it.

The divergence of lateral compressibility implies that extra
surface area can be pulled into the boundary region without
resistance. Thus, one expects the monolayer to attain a folded
structure that will be stabilized by higher, nonlinear-elastic
terms. A detailed description of this folded shape is beyond the
scope of the current work. Moreover, the transition from an
inflected to a folded state is treated here as a spinodal-like
instability. A more detailed study might yield a folded structure
as energetically favorable prior to this instability, \ie a
binodal-like transition preceding the one treated here. In a
macroscopic measurement of a pressure--area isotherm the
instability will appear as a plateau, much like any other
1st-order transition. Hence, distinguishing it from other
mechanisms requires microscopic visualization of the monolayer
\cite{ft_finite}.


\section{Diffuse domain boundary}
\label{sec_diffuse}

We now examine the effect of a domain boundary of finite
thickness.
That is, we suppose that rather than changing abruptly, $c_0$
varies gradually from $c_{01}$ to $c_{02}$ over a distance $\xi$.
We begin with a heuristic argument for the limit
of a very diffuse boundary, where $\xi$ is much larger than
$\lambda$ and $(\dc)^{-1}$. Consider a small element $\rmd s$ in
the boundary region. The difference in bending moments acting on
its two ends,
$K\rmd(\dot{\theta}-c_0)=K(\ddot{\theta}-\dot{c}_0)\rmd s$, is
balanced by a difference in tensile torque $\gamma\rmd
h=\gamma\sin\theta\rmd s$. In the limit of very large $\xi$ the
curvature changes very slowly, such that
$\ddot{\theta}\sim\theta_0/\xi^2$ is negligible compared to
$\dot{c}_0\sim -\dc/\xi$. We thus obtain
\begin{equation}
  \sin\theta_0 \sim {\lambda^2\dc}/{\xi} = {h}/{\xi}.
\label{t0_diffuse0}
\end{equation}
The mesa slope in this diffuse limit, unsurprisingly,
depends on $h/\xi$ rather than $h/\lambda$.
We can infer from Eq.~(\ref{t0_diffuse0}) several
less expected results as well.
The equation has no solution for $\theta_0>\theta_{0\rmc}=\pi/2$
(rather than $\pi$),
whereupon the mesa height is of the order of $\xi$,
\ie very large compared to $\lambda$.
Thus, at the critical compression the entire wide boundary lies
almost vertically, having a small lateral dimension.
The critical tension is
\begin{equation}
  \gamma_\rmc \sim K\dc/\xi,
\label{gc_diffuse0}
\end{equation}
\ie much smaller than its value in the sharp limit
($\sim K(\dc)^2$), implying that the monolayer can sustain much higher
compression than in the sharp case.
(Naturally, as $\xi$ tends to infinity one
expects the resulting almost-uniform monolayer to become increasingly
stable to the heterogeneity-driven folding.)

In order to check these results in more detail and study the
crossover between the sharp and diffuse limits, we now turn to
a detailed treatment of a simple, specific example.
Let us assume that the spontaneous
curvature $c_0$ changes linearly across the boundary,
\[
  c_0 = \left\{ \begin{array}{ll}
  c_{01} & s<-\xi/2 \\
  c_{01} - \dc (s/\xi+1/2) \ \ \ \ & -\xi/2<s<\xi/2 \\
  c_{02} & s>\xi/2.
  \end{array} \right.
\]
Substituting the spatial dependence of $c_0$
in the energy functional (\ref{g_general}) and taking the variation with
respect to $\theta(s)$, we obtain the following profile equations:
\begin{equation}
  \ddot\theta = \left\{ \begin{array}{ll}
  \lambda^{-2}\sin\theta & |s|>\xi/2 \\
  \lambda^{-2}\sin\theta - \dc/\xi \ \ \ & |s|<\xi/2.
  \end{array} \right.
\label{prof_diffuse}
\end{equation}
The boundary conditions in the current case are continuity of
$\theta$ and $\dot\theta$ at $s=\pm\xi/2$ and, as before, flatness
at infinity.

First integration of Eq.~(\ref{prof_diffuse}) gives
\begin{equation}
  \dot\theta^2 = \left\{ \begin{array}{ll}
  4\lambda^{-2} \sin^2\frac{\theta}{2} & |s|>\xi/2 \\
  4\lambda^{-2} \sin^2\frac{\theta}{2} - (2\dc/\xi)
  (\theta-\theta_1) & |s|<\xi/2,
  \end{array} \right.
\label{tdot_diffuse}
\end{equation}
where $\theta_1\equiv\theta(-\xi/2)=\theta(\xi/2)$. As in
Sec.~\ref{sec_sharp}, we can now calculate the energy of the
inflected conformation as a function of $\theta_0$ and $\theta_1$
by substituting the profile (\ref{tdot_diffuse}) back in the
energy functional (\ref{g_general}). The result is
\begin{equation}
  \frac{g(\theta_0,\theta_1)}{K\dc} = \frac{4}{\lambda\dc} \left\{
  4\sin^2\frac{\theta_1}{4} + \int_{\theta_1}^{\theta_0}\rmd\theta
  \left[ \sin^2\frac{\theta}{2} - \frac{\lambda^2\dc}{2\xi}
  (\theta-\theta_1) \right]^{1/2} \right\} - \theta_1.
\label{g_diffuse}
\end{equation}

Second integration, which would yield the topographic profile,
cannot be analytically performed in the current case. Instead, we
seek the equation for $\theta_0$ which replaces
Eq.~(\ref{t0_sharp}) for finding the point of instability.
Minimizing $g$ of Eq.~(\ref{g_diffuse}) with respect
to $\theta_0$ gives
one relation between $\theta_0$ and $\theta_1$,
\begin{equation}
  \theta_0-\theta_1 = \frac{2\xi}{\lambda^2\dc}
  \sin^2\frac{\theta_0}{2}.
\label{t0_diffuse1}
\end{equation}
This equation reflects a moment balance for the section 
$-\infty<s<0$; it is also obtained by setting $\dot\theta(0)=0$
in Eq.~(\ref{tdot_diffuse}).
Minimization with respect to $\theta_1$ yields
the second relation between $\theta_0$ and $\theta_1$,
%
%
%
\begin{equation}
  \int_{\theta_1}^{\theta_0} \left[ \frac{2\xi}{\lambda^2\dc}
  \sin^2\frac{\theta}{2} - (\theta-\theta_1) \right]^{-1/2} \rmd\theta
  = \left(\frac{\xi\dc}{2}\right)^{1/2}.
\label{t0_diffuse2}
\end{equation}
It stems from the moment balance on the section $-\xi/2<s<0$ and can
be also obtained from integration of 
the profile equation (\ref{prof_diffuse}) along this section.

\begin{figure}
\centerline{\epsfxsize=0.5\textwidth \hbox{\epsffile{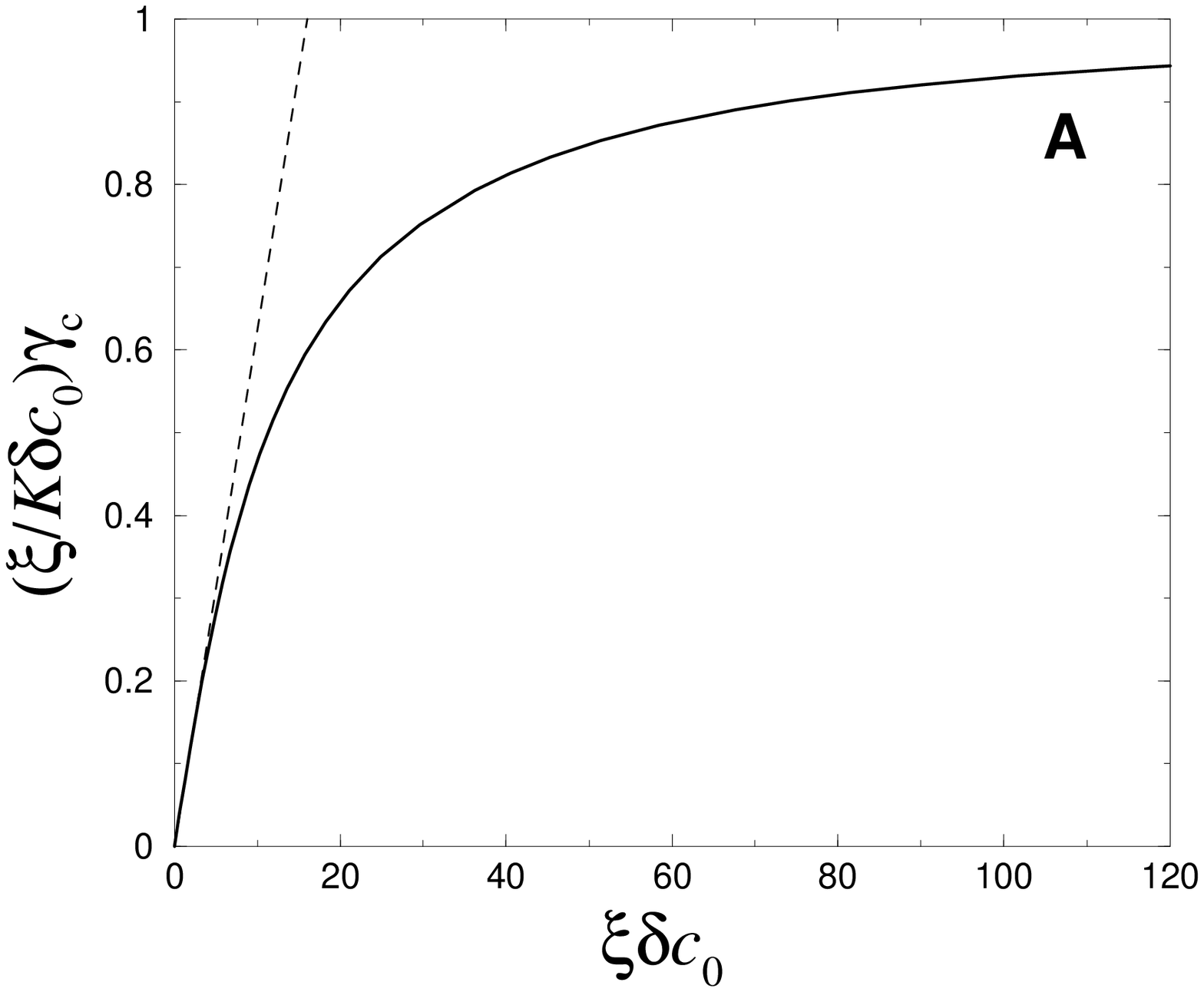}}
\epsfxsize=0.5\textwidth \hbox{\epsffile{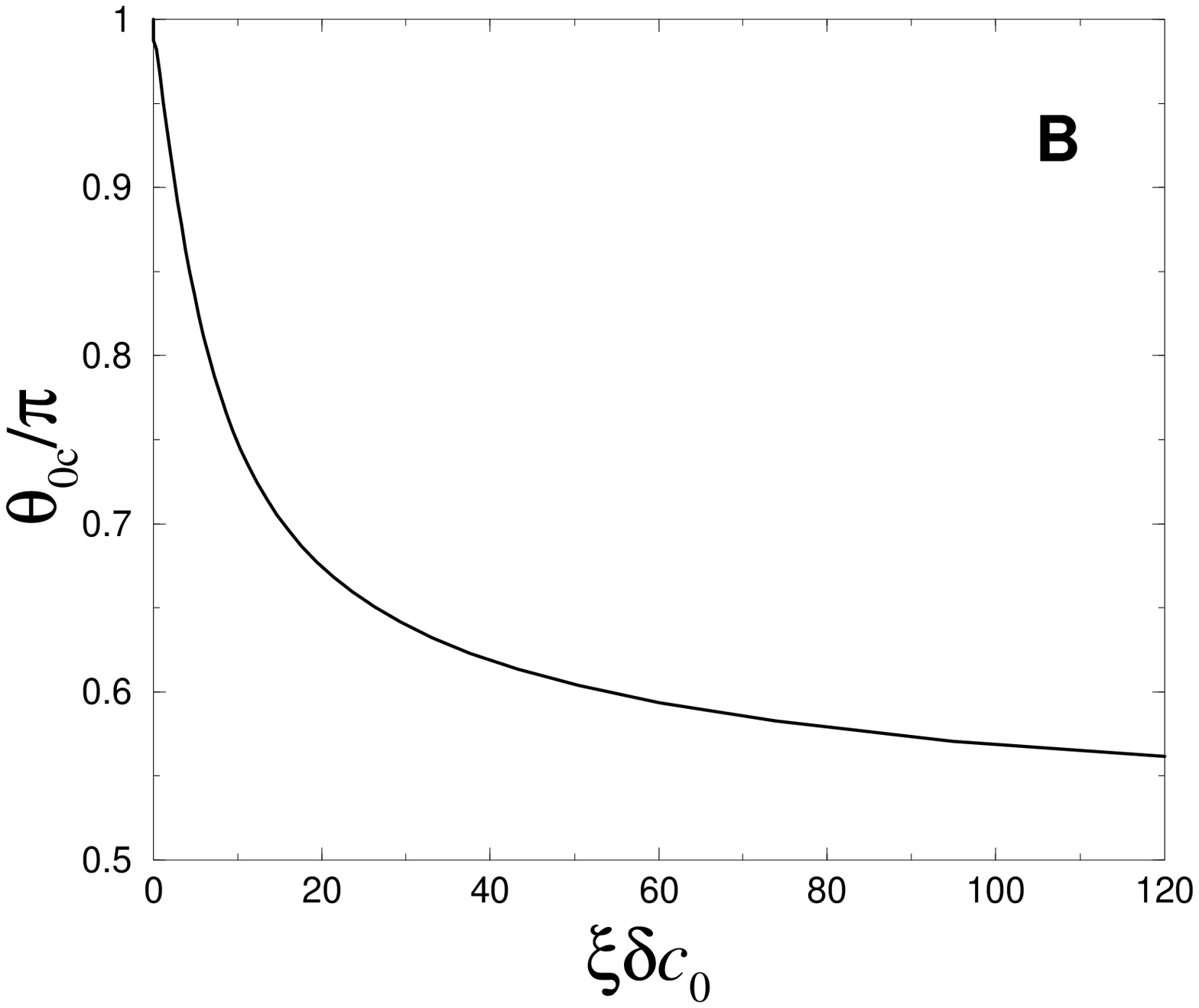}}}
\caption[]{Effect of diffuse boundary on mesa instability. 
A) Rescaled critical tension, 
$(\xi/K\dc)\gamma_\rmc$,
required for the transition.
The solid line is the result
for a boundary of finite thickness (\ie the scaling function $F(x)$ of
Eq.~(\ref{critical_diffuse2})). The dashed line is the
corresponding result for an infinitely sharp boundary (Eq.\
(\ref{critical_sharp})). B) Critical inflection
angle. For very sharp boundaries the angle at instability is
$\pi$, whereas for very diffuse ones it is reduced to $\pi/2$.} 
\label{fig_Fx}
\end{figure}

Equations (\ref{t0_diffuse1}) and (\ref{t0_diffuse2}) can be
solved numerically for $\theta_0$. As in Sec.~\ref{sec_sharp}, one
finds a minimum value $\gamma=\gamma_\rmc$ (\ie a maximum value of
$\lambda$) beyond which there is no solution to the equations and
the monolayer becomes unstable. From the fact that the various
parameters are grouped in Eqs.\ (\ref{t0_diffuse1}) and
(\ref{t0_diffuse2}) into two dimensionless terms, $\xi\dc$ and
$\xi/(\lambda^2\dc)=\xi/h$, it follows that the critical tension
must satisfy the scaling law
\begin{equation}
  \gamma_\rmc = \frac{K\dc}{\xi} F(\xi\dc)
\label{critical_diffuse}
\end{equation}
where $F(x)$ is a certain ``universal", dimensionless function. In
Fig.~\ref{fig_Fx}A we have plotted the function $F(x)$ as obtained
from numerical integration of Eqs.\ (\ref{t0_diffuse1}) and
(\ref{t0_diffuse2}).

In the sharp limit, $\xi\dc\ll 1$, $\gamma_\rmc$ must become
independent of $\xi$. Hence, $F(x)$ is linear for small $x$ so as
to get $\gamma_\rmc\sim K(\dc)^2$.
Indeed, in this limit Eq.~(\ref{t0_diffuse2}) reduces to
$\theta_0-\theta_1\simeq\xi\dc/8$ which, together with
Eq.~(\ref{t0_diffuse1}), recovers the results of Sec.~\ref{sec_sharp}
(cf.\ Eqs.\ (\ref{t0_sharp}) and (\ref{critical_sharp})),
  $\sin(\theta_0/2) \simeq \lambda\dc/4=h/(4\lambda)$,
  $\theta_{0\rmc} \simeq \pi$,
  and $\gamma_\rmc \simeq \frac{1}{16} K(\dc)^2$.
We thus conclude that the results of Sec.~\ref{sec_sharp} are
valid as long as $\xi\ll (\dc)^{-1}$.

In the diffuse limit, $\xi\dc\gg 1$, Eq.~(\ref{t0_diffuse2})
reduces to
\begin{equation}
  \sin\theta_0 \simeq \lambda^2\dc/\xi=h/\xi,
\label{t0_diffuse3}
\end{equation}
as was anticipated in Eq.~(\ref{t0_diffuse0}).
This leads to
 $\theta_{0\rmc} \simeq \pi/2$,
 $\gamma_\rmc \simeq {K\dc}/{\xi}$.
Hence $F(x)\simeq 1$ for $x\gg 1$, as is verified in Fig.~\ref{fig_Fx}A.

To summarize these results:
\begin{eqnarray}
  \gamma_\rmc &=& \frac{K\dc}{\xi} F(\xi\dc) \nonumber\\
  F(x) &\simeq& \left\{ \begin{array}{ll}
  x/16 \ \ \ \  & x\ll 1\\
  1 & x\gg 1.
  \end{array}\right.
\label{critical_diffuse2}
\end{eqnarray}
Recall that we have obtained the results in both the sharp and diffuse
limits independently of the detailed shape of $c_0(s)$. Hence,
varying the spatial dependence of $c_0$ would merely affect the
exact shape of $F(x)$ in between these limits. The more diffuse
the boundary, the higher the compression required for folding, the
higher the mesa wall at the instability, and the smaller the
critical inflection angle. Diffuse boundaries thus allow a
biphasic monolayer to withstand stronger compression and higher
mesas. Despite the smaller inflection angle there is always an
overhang topography at the instability, \ie
$\pi/2<\theta_{0\rmc}<\pi$. The dependence of the critical
inflection angle on boundary thickness is shown in
Fig.~\ref{fig_Fx}B.


\section{Instability of the domain boundary}
\label{sec_rippling}

\begin{figure}
\centerline{\epsfxsize=0.55\textwidth
\hbox{\epsffile{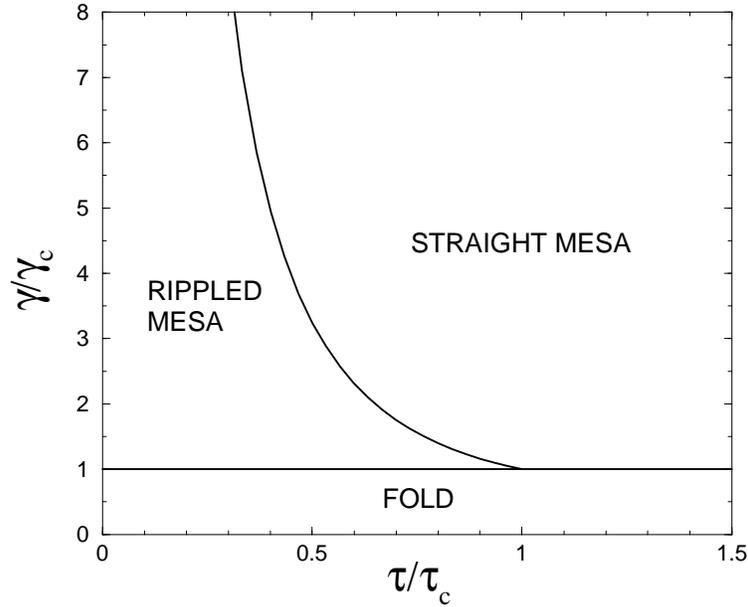}}}
\caption[]{Surface
tension--line tension diagram of topographies for a sharp domain
boundary. If $\tau>\tau_\rmc=(\pi-2)K\dc$ the mesa topography
remains straight upon decreasing surface tension until it becomes
unstable at $\gamma=\gamma_\rmc=\frac{1}{16}K(\dc)^2$. If
$\tau<\tau_\rmc$ the mesa wall ripples below a surface tension
$\gamma=\gamma_\rmr>\gamma_\rmc$.} \label{fig_ripplediag}
\end{figure}

Until now we have considered only topographies which do not vary
along the direction of the domain boundary, and thus do not affect
its length. The departure from a flat conformation near a domain
boundary, as studied in the previous sections, is energetically
beneficial, \ie the inflection energy per unit length, $g$, is
negative. Hence, as far as the topographic effect is concerned, it
would be favorable to increase the boundary length. In other
words, the topography effectively reduces the line tension of the
phase boundary, the reduction being given by $g$ of
Eqs.~(\ref{g_sharp2}) or (\ref{g_diffuse}) for a sharp or diffuse
boundary, respectively. Consequently, if the bare line tension of
the boundary \cite{ft_linetension},
$\tau$, is smaller than the maximum value of $|g|$,
then, for a certain inflection angle $\theta_0<\theta_{0\rmc}$,
the effective line tension will turn negative and one expects the
boundary to ripple. Assuming hereafter a sharp boundary, we obtain
the condition for rippling by setting
$\theta_0=\theta_{0\rmc}=\pi$ in Eq.~(\ref{g_sharp2})
\cite{ft_nl2},
\begin{equation}
  \tau < \tau_\rmc = (\pi-2)K\dc.
\label{tau_c}
\end{equation}
The inflection angle and surface tension
required for rippling, $\theta_{0\rmr}$ and
$\gamma_\rmr$, are then obtained from the equations
\begin{eqnarray}
  && \theta_{0\rmr} - 2\tan\frac{\theta_{0\rmr}}{4} =
  \frac{\tau}{K\dc} \nonumber\\
  && \gamma_\rmr = \frac{\gamma_\rmc}{\sin^2(\theta_{0\rmr}/2)},
\label{theta_r}
\end{eqnarray}
where, as defined in Eq.~(\ref{critical_sharp}),
$\gamma_\rmc=\frac{1}{16}K(\dc)^2$.
The diagram in Fig.~\ref{fig_ripplediag} summarizes the results
concerning the topographic
transitions near a sharp boundary as a function of surface tension
and line tension.

\begin{figure}
\centerline{\epsfxsize=\textwidth
\hbox{\epsffile{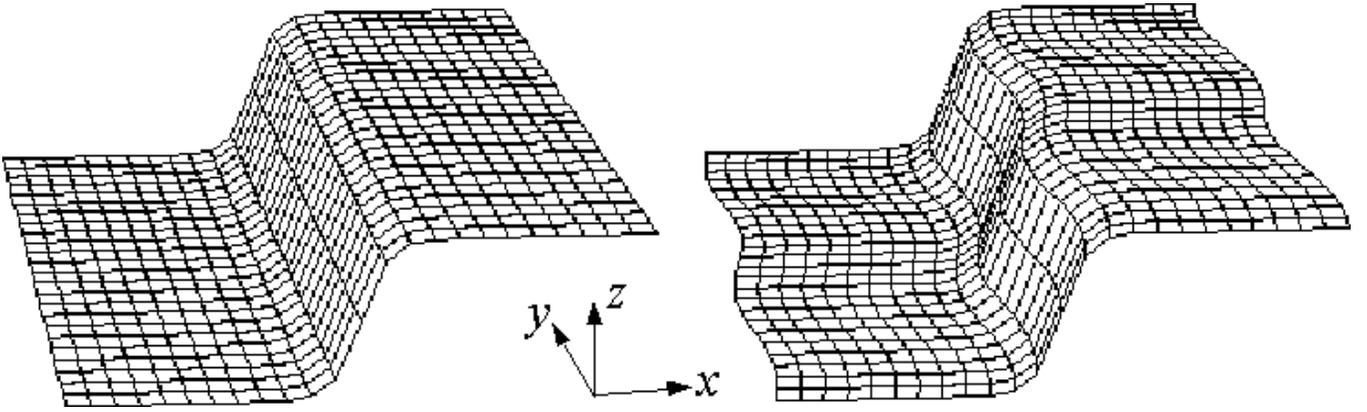}}}
\vspace{.5cm}
\caption[]{Schematic sketch of the assumed rippling perturbation.}
\label{fig_ripple}
\end{figure}

Let us now examine the spatial form of the rippling transition. In
Sec.~\ref{sec_sharp} we assumed a straight, sharp domain boundary,
which can be represented in Cartesian coordinates as the line
$\vecR(y)=(x=0,y,z=0)$. We now wish to perturb the inflected
conformation by considering a boundary that slightly wiggles with
amplitude $a$ and wave number $q$. The full three-dimensional
problem is formidable. We therefore restrict ourselves to a simple
subset of perturbations---uniform displacements of the inflected
shape in the $x$ direction, whose magnitude undulates in the $y$
direction (see Fig.~\ref{fig_ripple}). Since we do not exhaust all
available conformations, the minimum energy that we are about to
calculate might be higher than the true minimum. Hence, the
following results should be considered as an upper-bound estimate
for the rippling instability. Nevertheless, this estimate is
expected to be good as long as the wiggling wavelength is much
larger than the inflection extent, $q\lambda\ll 1$. In this limit
the two lateral length scales can be separated, as has been done in
Sec.~\ref{sec_sharp}, and one expects the preferred perturbations
to resemble that of Fig.~\ref{fig_ripple}. Employing this
simplification, we can represent the perturbed boundary by the
curve $\vecR(y)=(x=a\sin qy,y,z=0)$, and conveniently parameterize
the monolayer surface as
\begin{eqnarray*}
   \vecr(s,t) &=& (x(s,t),y(s,t),z(s,t)) \\
   x(s,t) &=& \int_0^s \cos\theta(s')\rmd s' + a\sin qt,\ \ \
   y(s,t) = t,\ \ \
   z(s,t) = \int_0^s \sin\theta(s')\rmd s',
\end{eqnarray*}
such that in the ``material coordinates" $(s,t)$ the boundary line
is again given by $\vecR(t)=(s=0,t,z=0)$.

In order to use the energy functional (\ref{G_general}) we need
to represent the geometrical parameters of the surface---its
mean curvature $c(s,t)$, area element $\rmd A(s,t)$,
projected area element $\rmd A_{\rm p}(s,t)$, and boundary
arc-length $\rmd R(s,t)$---using the new coordinates.
This technical calculation is presented in the Appendix.

Substituting Eqs.~(\ref{dA})--(\ref{cst}) into the
elastic energy expression, Eq.~(\ref{G_general}), and expanding to
second order in the rippling amplitude $a$, we obtain
\begin{eqnarray}
  g[\theta(s)] &\equiv& G/L = g^{(0)} + a^2 g^{(1)} \nonumber\\
  g^{(1)} &=& \frac{1}{4} \left( \tau' q^2 + b q^4 \right),
\label{g_ripple}
\end{eqnarray}
where $g^{(0)}[\theta(s)]$ is the
energy functional in the straight case, given by
Eq.~(\ref{g_general}), and $L=\int\rmd t$.
The coefficients $\tau'$ and $b$ are functionals of the topography
$\theta(s)$:
\begin{eqnarray}
  \tau' &=& \tau +
  \int\rmd s \left[ K \left( 2-\frac{5}{2}\sin^2\theta
  \right)\dot{\theta}^2 - 2Kc_0\cos^2\theta\dot{\theta} +
  \gamma\sin^2\theta \right]  \nonumber\\
  b &=& K\int\rmd s\sin^2\theta.
\label{taub}
\end{eqnarray}
They act as effective line tension and bending modulus,
respectively. The one-dimensional bending modulus $b$ is
proportional to $K$, the two-dimensional modulus of the sheet;
when the boundary curves the monolayer must bend with it
(see Fig.~\ref{fig_ripple}), leading to a cost in bending energy.

In principle one could now minimize $g$ of Eq.~(\ref{g_ripple})
with respect to $\theta(s)$ and find the perturbed shape,
$\theta(s)=\theta^{(0)}(s)+a^2\theta^{(1)}(s)$. However, since
$\delta g^{(0)}/\delta\theta^{(0)}=0$, substituting the perturbed
shape back in $g$ would yield, up to order $a^2$,
$g=g^{(0)}[\theta^{(0)}]+a^2 g^{(1)}[\theta^{(0)}]$. Thus, if we
are merely interested in the perturbed {\it energy}, we may just
substitute in Eq.~(\ref{taub}) the unperturbed topography
$\theta^{(0)}(s)$ as found in Sec.~\ref{sec_sharp}
(Eqs.~(\ref{tdot_sharp}) and (\ref{t0_sharp})). This yields
\begin{eqnarray}
  \tau' &=& \tau - K\dc \left(
  \theta_0 - 2\tan\frac{\theta_0}{4} \right) \nonumber\\
  b &=& \frac{32K}{3\dc} \sin\frac{\theta_0}{2}
  \left( 1-\cos^3\frac{\theta_0}{2} \right).
\label{taup_b}
\end{eqnarray}

When the effective line tension vanishes, $\tau'=0$, there is a
$q=0$ (\ie 2nd-order) rippling transition, as already anticipated
in Eqs.~(\ref{tau_c}) and (\ref{theta_r}). (Strictly speaking,
since the domain boundary is finite and closed, the transition is
encountered only at $\tau'=-\pi^2b/L^2$, \ie for the lowest-order
undulation of $q=\pi/L$.) The rippling of the one-dimensional
boundary is thus analogous to the Euler buckling of an elastic rod
\cite{LL}.

\begin{figure}
\centerline{\epsfxsize=0.63\textwidth \hbox{\epsffile{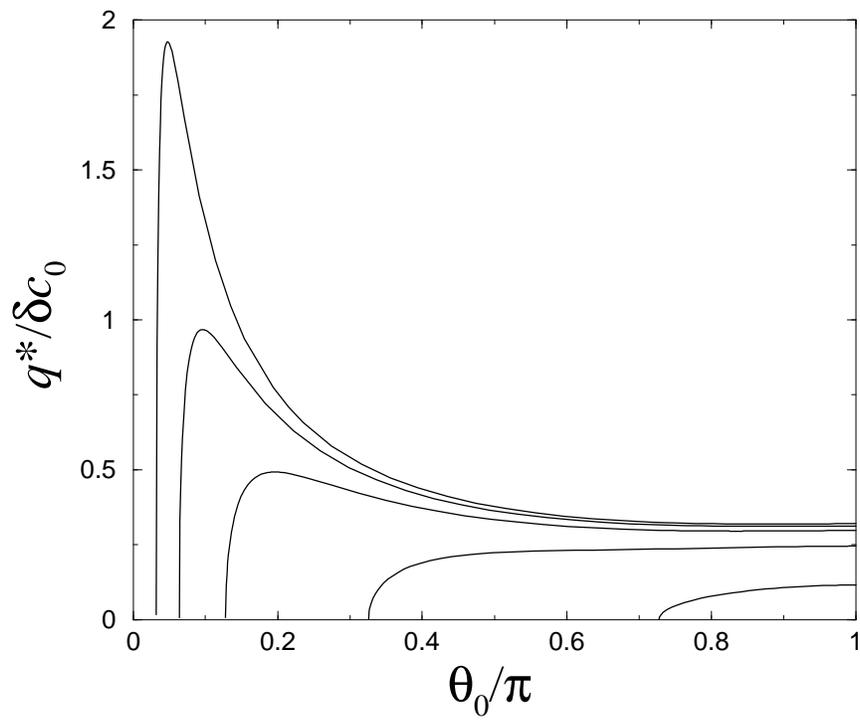}}}
\caption[]{Dependence of rippling wavenumber on inflection angle
for various values of line tension (from top to bottom):
$\tau/(K\dc)=0.05, 0.1, 0.2, 0.5, 1$. For higher line tension,
$\tau/(K\dc)>\pi-2\simeq 1.14$, there is no rippling.}
\label{fig_qstar}
\end{figure}

Upon further compression, or if the monolayer is ``quenched" to
$\tau'<0$, all modes satisfying
\begin{equation}
  q < q^* = \left(\frac{-\tau'}{b}\right)^{1/2} =
  \frac{\sqrt{3}}{4\sqrt{2}} \dc \left[
  \frac {\theta_0-2\tan(\theta_0/4) - \tau/(K\dc)}
  {\sin(\theta_0/2)[1-\cos^3(\theta_0/2)]} \right]^{1/2},
\label{qstar}
\end{equation}
become unstable and their amplitudes start growing.
We expect the observed unstable modes to have roughly the same
scale as the upper bound $q^*$. The scale of the rippling wavelength
is thus set by $(\dc)^{-1}$, which is usually much smaller than 
the boundary length $L$. Hence,
although this is strictly a $q=0$ instability, one expects
in practice to observe a densely wiggling boundary on the scale of
the entire domain.
Another interesting observation is that, beyond the 
onset of rippling, $q^*$ does not always increase
monotonically with compression. The nonmonotonic behavior becomes
more pronounced the smaller the value of $\tau/(K\dc)$, as
demonstrated in Fig.~\ref{fig_qstar}. For small values of this
parameter, therefore, one expects the boundary to ripple densely
beyond the onset of instability and then, upon further
compression, return to a less rough shape.
Recall that our Ansatz concerning the preferred perturbation 
is expected to give reliable results as long as 
$q\ll\lambda^{-1}$. We have found that the rippling modes obey
$q\lesssim\dc$. On the other hand, 
a stable, sharp topography requires $\lambda^{-1}\gtrsim\dc/4$
(see Eq.~(\ref{t0_sharp})).
Thus, our assumption is only marginally fulfilled and the results
should be regarded merely as a qualitative guide.


\section{Behavior near a critical point}
\label{sec_Tc}

The topographic effects described in this article rely on a
contrast between different domains. Hence, when a
monolayer at coexistence reaches a critical point, these
effects are expected to vanish along with the domain structure.
Various parameters affecting the topography dramatically change
when the critical point is approached: the density contrast
becomes increasingly weak (leading to a smaller $\dc$), domain
boundaries get diffuse (larger $\xi$), and the bare
line tension between domains, $\tau$, tends to zero. Thus,
although the topography must clearly disappear at the critical
point, its exact behavior has to be examined in detail. For
example, it is unclear {\it a priori} whether, with respect to
topography, the monolayer is driven towards the diffuse limit
(larger $\xi\dc$) or the sharp one (smaller $\xi\dc$).

As the temperature $T$ approaches its critical value $T_\rmc$,
$\hat{T}\equiv|T-T_\rmc|/T_\rmc\rightarrow 0$, we have \cite{CP}
\begin{eqnarray}
  \xi &\sim& \hat{T}^{-\nu} \rightarrow \infty \nonumber\\
  \dc &\sim& \hat{T}^\beta \rightarrow 0,
\end{eqnarray}
where, for a two-dimensional fluid, $\nu=1$ and $\beta=1/8$.
%
Hence, the height difference, given by Eq.~(\ref{sumrule}), decays
as
\begin{equation}
  h = \lambda^2\dc \sim \hat{T}^\beta = \hat{T}^{1/8}.
\end{equation}
Since $\xi\dc\sim \hat{T}^{-\nu+\beta}\sim
\hat{T}^{-7/8}\rightarrow\infty$, it is the diffuse limit of
Sec.~\ref{sec_diffuse} that applies near the critical point.
(The three-dimensional topography could affect the critical
behavior of the two-dimensional fluid in the monolayer as a
``hidden", annealed variable. Hence, the critical exponents should
be modified according to the Fisher renormalization \cite{Fisher}.
In the case of a two-dimensional fluid (or Ising model), however,
the Fisher renormalization leaves the exponents intact.)

We now explore further details of the topographic critical
behavior. For small inflection angles we expect in the diffuse
limit (cf.\ Eq.~(\ref{t0_diffuse3}))
\begin{equation}
  \theta_0 \simeq h/\xi \sim t^{\nu+\beta} = \hat{T}^{9/8}.
\label{t0_Tc}
\end{equation}
Indeed, the topography has been found in Sec.~\ref{sec_diffuse} to
depend on two dimensionless quantities, $\xi\dc$ and
$\xi/(\lambda^2\dc)=\xi/h$, both of which diverge at the critical
point---the former as $\hat{T}^{-\nu+\beta}=\hat{T}^{-7/8}$
and the latter as
$\hat{T}^{-\nu-\beta}=\hat{T}^{-9/8}$.
Studying Eqs.\ (\ref{t0_diffuse1}) and
(\ref{t0_diffuse2}) in this asymptotic limit, one finds
$\theta_0\simeq h/\xi$ and $\theta_1\simeq\theta_0/2$, which
verifies Eq.~(\ref{t0_Tc}). Substituting these results in
Eq.~(\ref{g_diffuse}) for the inflection energy, we get
\begin{equation}
  g \simeq -\frac{K\lambda^2\dc^2}{2\xi}
  \sim \hat{T}^{\nu+2\beta} = \hat{T}^{5/4}.
\label{g_Tc}
\end{equation}
Thus, the topographic contribution to the heat capacity of the
monolayer vanishes as 
$\partial g/\partial\hat{T}\sim\hat{T}^{1/4}$, 
whereas the heat capacity
of the two-dimensional fluid diverges logarithmically \cite{CP}.
This consistently demonstrates that the critical behavior of the
monolayer remains unaffected by the topography.

How does the approach to the critical point influence the
instabilities studied in the previous sections? The folding
instability in the diffuse limit requires, according to
Eq.~(\ref{critical_diffuse2}), a surface tension lower than
\begin{equation}
  \gamma_\rmc\simeq K\dc/\xi \sim \hat{T}^{\nu+\beta}
  \sim \hat{T}^{9/8} \rightarrow 0.
\end{equation}
In practice, therefore, as soon as the required lateral pressure
exceeds that of the critical point the folding instability will
become unattainable.

The effect on the boundary-rippling instability studied in
Sec.~\ref{sec_rippling} is more delicate. Upon approaching the critical
point the bare line tension of the boundary gets vanishingly small
as \cite{Onsager,Widom}
\begin{equation}
  \tau \sim \hat{T}^\mu \rightarrow 0,\ \ \ \mu=1.
\end{equation}
Thus, the resistance to rippling becomes increasingly weak. Yet,
at the same time the driving force for rippling, \ie the energy
$g$ gained due to the inflected topography, gets weaker as well.
According to Eq.~(\ref{g_Tc}) the latter vanishes slightly faster,
as $\hat{T}^{5/4}$. Hence, it is the bare line tension that wins
close to the critical point, and the boundary
topography should flatten out at $T_\rmc$ as a smooth step without
ripples.


\section{Discussion}
\label{sec_discussion}

We have demonstrated in this work that biphasic monolayers are
generally nonflat, having inflected shapes in the vicinity of
domain boundaries. This leads to an overall topography of mesas
where domains of one phase are higher than those of the other. As
the monolayer is progressively compressed the mesas grow more
pronounced, subsequently developing overhangs, and finally
becoming unstable.

Substituting typical values for phospholipid monolayers
\cite{Safran}---$\gamma\simeq$ 10--50 erg/cm$^2$, $K\simeq$ 10--50
$k_{\rm B}T$, $c_0^{-1}\simeq$ 5--10 nm---we get $\lambda\simeq$
1--10 nm, $\lambda\dc\simeq$ 0.1--1, and $h\simeq$ 0.1--10 nm.
Hence, the mesas are steep but low. The numerical value of
$\lambda\dc$ implies that the predicted instability
($\lambda\dc\geq 4$) may be observed for attainable pressures. The
energy per unit length gained by departing from the flat state to
a sharp inflection is, according to Eq.~(\ref{g_sharp2}), $g\simeq
K\dc\simeq$ 1--10 $k_{\rm B}T$/nm. Hence, for a typical domain
size of $L\sim$ 1--10 $\mu$m, the inflected conformation is
``frozen", \ie robust under thermal fluctuations. This justifies
our mechanical, ``zero-temperature" approach.

As a more specific example, we may consider a monolayer whose
behavior is governed by electrostatic interactions. The deviation
from a flat conformation is thereby driven by variations in the
lateral charge density, $\sigma(s)$. In the typical case of strong
screening, $c/\kappa\ll 1$, where $\kappa^{-1}$ is the Debye
screening length, one obtains \cite{Glen}
$Kc_0={\pi\sigma^2}/{\eps\kappa^2}$ and
$K={3\pi\sigma^2}/{2\eps\kappa^3}$, $\eps$ being the dielectric
constant of water. (Note the finite, positive $c_0$; charged
monolayers spontaneously tend to curve into the aqueous phase.)
Consequently, substituting typical values of $\sigma\simeq$ 1
charge per 0.3--1 nm$^2$ and $\kappa^{-1}\simeq$ 1--10 nm, we
reach similar conclusions to those above.

We have studied domain boundaries of finite thickness as well. The
qualitative features of inflected conformation and instability do
not disappear for any boundary thickness $\xi$. On one hand, for a
given compression a diffuse boundary leads to more
moderate slopes compared to a sharp one.
On the other hand, it shifts the folding
instability to a higher pressure, thus strengthening the monolayer
and allowing for higher mesas to be stabilized. Unfortunately,
conventional means of increasing $\xi$, \eg heating towards a
critical point, also reduce the domain contrast $\dc$, thus
suppressing the topography. We have studied this delicate
interplay close to a critical point in Sec.~\ref{sec_Tc}. Our
results for diffuse boundaries show that the simple, infinitely
sharp limit gives good results as long as $\xi<(\dc)^{-1}$, which
holds in most practical circumstances except near a critical
point.

One might worry about additional factors that would destroy the
inferred topography. Such a factor is the cost in gravitational
energy of displacing water from the flat interface. This energy
per unit area is about $\delta\rho gh^2 \sim 10^4$ $k_{\rm
B}T/\mbox{cm}^2$, where $\delta\rho$ is the difference in density
of the two phases and $g$ is here the gravitational acceleration. 
Thus, due to the small height of the mesas
(1--10 nm), gravity is negligible over all relevant lateral length
scales (up to meters). (Beyond the topographic instability,
however, the monolayer may become much more folded, and gravity
may have a significant stabilizing role.) Another factor to worry
about is the van der Waals attraction between the inferred
overhang and the underlying surface, which might make the overhang
collapse. The attraction energy per unit area is roughly $H/h^2$,
where $H$ is the Hamaker constant divided by $12\pi$ (typically a
few $k_{\rm B}T$) and $h\simeq\lambda^2\dc$ is the overhang height
\cite{Israelachvili}. The lateral extent of the overhang is
$\lambda$, and the resulting energy per unit length,
$H/(\lambda^3\dc^2)$, is to be compared with the inflection
energy, $K\dc$. The ratio is $(H/K)(\lambda\dc)^{-3}\ll 1$, since
$K$ of a lipid monolayer is a few tens $k_{\rm B}T$ and
$\lambda\dc\simeq$ 3--4 to get an overhang (cf.\
Eq.~(\ref{t0_sharp})). Hence, the van der Waals attraction is too
weak to significantly affect the overhang.

The topography of mesas and overhangs is thus a robust result
which should be observable in practice. Such an observation is
difficult, however, because of the small height differences and
fluidity of the interface. Very recently a new experimental
technique has been presented, utilizing non-specular scattering of
intense light to visualize small topographic features in
phopholipid monolayers \cite{Vogel}. Although the study was
focused on features of a pure liquid-condensed phase, height
differences were reported at boundaries of liquid-expanded domains
coexisting with a gas phase, as well as liquid-condensed domains
in a liquid-expanded phase. (Interestingly, a stronger signal was
obtained in the former case, perhaps due to a larger contrast in
spontaneous curvature.) It is still unclear whether these 
experimental findings are related to the topography discussed here
or to other, more molecular effects.

\begin{figure}
\centerline{\epsfxsize=0.5\textwidth
\hbox{\epsffile{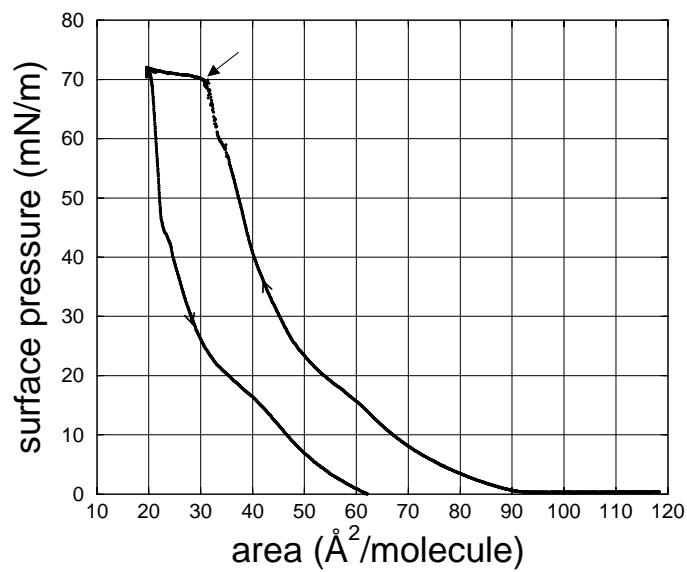}}}
\caption[]{Pressure--area isotherm for a mixed monolayer of DPPC
and POPG, as measured during a compression/expansion cycle in a
Langmuir trough. The mole ratio is DPPC:POPG$=$7:3 and the
temperature 25$^\circ$C. The folding instability is indicated by
an arrow.} \label{fig_iso}
\end{figure}

Recent experiments on mixed phospholipid monolayers have revealed
a new type of folding instability \cite{KaYee,ourEPL}. When the
monolayer is compressed and enters a coexistence region, there
is a critical pressure at which micron-scale folds appear. The
folding is significantly more reversible than other collapse
mechanisms and is therefore thought to be of key importance to the
function of lungs. Figure~\ref{fig_iso} shows a pressure--area
isotherm as measured for a mixed phospholipid monolayer of
dipalmitoylphosphatidylcholine (DPPC) and
palmitoyloleoylphosphatidylglycerol (POPG). The folding is
manifested by a plateau in the isotherm. (The same phenomenon was
observed in DPPG monolayers at a much higher surface tension
\cite{KaYee}.) Figure~\ref{fig_fold} presents a sequence of
fluorescence microscopy images of the monolayer just before and
just after the instability.

We believe that this folding phenomenon is initiated by the
topographic instability of boundary regions as obtained from our
model. (Further evolution and propagation of the fold are
determined by other factors not taken into account in the current
work, such as the viscoelasticity of the monolayer
\cite{ourEPL,visco}.) If the hypotheses regarding the biological
significance of the folding and its relation to topography are
correct, it may represent an interesting solution of Nature to a
delicate mechanical problem. Using a mixed surfactant monolayer to
cover the lung leads to domain formation upon compression, which
in turn allows the topographic instability and folding. Additional
constituents (\eg proteins) may ensure that the folding is not
preceded by other, irreversible collapse mechanisms \cite{KaYee}.
This design provides the monolayer with a unique way to yield
gracefully to compression and reduce its projected area, while 
avoiding irreversibility and loss of surfactant.

Folding of the mesa structure is in many cases preempted by other
instabilities. One type of collapse is delamination---breakage of
the monolayer into multiple layers \cite{multilayer1,multilayer2}. 
It occurs
when the surfactant sheet yields to a combination of bending and
lateral compressive stresses. Since the mesas help relieve part of
the inherent bending stresses exerted in a flat monolayer, one
expects the breakage to occur (somewhat counter-intuitively) away
from the boundary, inside the more frustrated domain (\ie the
one having higher spontaneous curvature). Another mode of
monolayer collapse is budding and ejection of vesicles into the
aqueous phase \cite{ejection}. Recent experiments on mixed
phospholipid monolayers have shown that vesiculation is promoted
by increased temperature and may coexist with folded structures
\cite{Ajay}. The effect of topography on delamination and budding,
as well as the interplay between the various collapse modes, are
yet to be studied in detail.

Another general conclusion arising from this work relates to fluid
surfaces of vanishing tension. Such surfactant films are
encountered, \eg in emulsions, $L_3$ (``sponge") phases and large,
unsupported bilayer vesicles \cite{Safran}. 
The topographic instability found
for a finite tension implies that these tensionless surfaces
cannot sustain a stable domain structure. To the best of our
knowledge a static domain structure has never been observed in
those systems. We attribute it to the inevitable shape instability
that would occur near domain boundaries if such a structure
existed.

We have studied a possible rippling of the domain boundary upon
compression. This phenomenon arises from a competition between the
topographic features accompanying the domain boundary and its bare
line tension. The threshold value of line tension required to get
rippling is $\tau_\rmc\simeq g\simeq$ 1--10 $k_{\rm B}T$/nm, which
is of the same order as line tension values measured in
experiments
\cite{Knobler,ltension1,ltension2,ltension3,ltension4}. The
rippling is therefore a realistic, observable feature. There is
already a well-established mechanism for shape transformations of
monolayer domains, driven by a competition between line tension
and long-range electrostatic interactions
\cite{McConnell,Seul,Goldstein,Knobler}. We offer the topographic
rippling as an additional mechanism which should be observed in
practice. There are three major features distinguishing the two
phenomena. (i) In the electrostatic mechanism an infinite straight
boundary is never stable. Consequently, the stable domain size and
wavelength of boundary instability have the same scale, $L\sim
q^{-1}\sim l\rme^{\tau/\delta p^2}$, where $l$ is a molecular size
and $\delta p$ the difference in dipole densities of the two
phases \cite{McConnell}. Hence, shape deformations occur on the
scale of the entire domain \cite{Knobler,McConnell,Seul}, leading
to a sequence of well-resolved, ``quantized" transitions. By
contrast, the length scale of the topographic rippling,
$(q^*)^{-1}\sim(\dc)^{-1}\lesssim 0.1$ $\mu$m, 
is much smaller than, and unrelated
to, $L$. 
Thus, we expect
this instability to appear as a small-scale roughening of the
domain boundary. (ii) The topographic rippling, being an elastic
mechanism, should not be very sensitive to changes in
electrostatic parameters such as ionic strength and molecular
charge. (iii) As demonstrated in Sec.~\ref{sec_rippling}, the
rippling wavelength may exhibit in certain circumstances a
peculiar nonmonotonic behavior as a function of pressure. In
addition, we have shown in Sec.~\ref{sec_Tc} that topographic
rippling is inhibited near a critical point. Hence, boundary
rippling could be smoothed out by heating the monolayer towards
its critical temperature. This is in contrast with common surface
instabilities that are usually promoted by increasing temperature.

In a recent experiment on a pure DPPC monolayer at liquid
expanded--liquid condensed coexistence, a domain-boundary
instability of sub-micron scale has been observed \cite{Canay}.
This is demonstrated in Fig.~\ref{fig_ripple_exp}. At a critical
pressure slight roughening appears simultaneously in all domain
boundaries. Upon little further compression the roughening becomes
denser and the boundaries look fuzzy due to optical limitations. A
detailed presentation of this effect will be given in a
forthcoming publication \cite{Canay}. The small length scale of
this instability (compared to the entire domain size) is in accord
with the topography-induced mechanism discussed above. Yet,
further study is required in order to clarify the relation between
the two effects.

The phenomena described in this work---mesa formation, mesa
instability, boundary rippling---arise from rather basic
considerations. Nevertheless, there is still a gap between theory
and experimental observations. The relation between the
topographic instability as obtained from the elastic model and the
observed folding in biphasic lipid monolayers is still to be
established. In particular, the current theory does not account
for the fully developed folded structure and its stability, as
observed in experiments. Topography-induced boundary rippling and
its distinction from the known electrostatic mechanism is yet
another intriguing feature to be experimentally investigated. We
hope to close this gap in future publications.

The mesa topography is a novel interfacial feature predicted 
by our work. If mesas exist, which is yet to be decisively proven 
by experiment, they should have important implications on 
various interfacial aspects, such as surface interactions with 
dissolved molecules, behavior in confined geometries, and 
possible applications for controllable nanostructures.
We would like to draw special attention to the unique overhang
topography predicted by the model. Under the right compression all
the domains in a biphasic monolayer should develop regular lips at
their edges. Such controllable nanoscale grooves might be
technologically useful, \eg for capturing and encapsulating
(bio)polymers.


\acknowledgments

This work was supported by the National Science Foundation
under Awards Nos.\ DMR 9975533 and 9728858, and by its 
MRSEC program under Award No.\ 9808595.
HD was partially supported by the American Lung Association
(RG-085-N).
CE was supported by the American Health Assistance Foundation
(A1999057) and the Altzheimer's Association (IIRG-9901175).
AG was supported by the Searle Scholars Program/The Chicago
Community Trust (99-C-105).
The experimental apparatus was made possible by an NSF 
CRIF/Junior Faculty Grant (CHE-9816513).
KYCL is grateful for support from the March of Dimes Basil
O'Connor Starter Scholar Research Award (5-FY98-0728), and
the David and Lucile Packard Foundation (99-1465).

\appendix

\section*{Differential geometry of a rippled boundary}

In the rippled state we represent the domain boundary by the curve
$\vecR(y)=(x=a\sin qy,y,z=0)$ and parameterize the monolayer
surface as
\begin{eqnarray}
   \vecr(s,t) &=& (x(s,t),y(s,t),z(s,t)) \nonumber\\
   x(s,t) &=& \int_0^s \cos\theta(s')\rmd s' + a\sin qt,\ \ \
   y(s,t) = t,\ \ \
   z(s,t) = \int_0^s \sin\theta(s')\rmd s'.
\label{param}
\end{eqnarray}
We now need to represent the various properties of the surface
using the ``material coordinates" $(s,t)$ \cite{DiffGeom}.

The determinant of the metric tensor associated with the surface 
of Eq.~(\ref{param}) is
\begin{equation}
  \Gamma = (\partial_s\vecr\times\partial_t\vecr)^2 =
  1 + (qa \cos qt \sin\theta)^2.
\end{equation}
The Jacobian of the transformation $(x,y)\rightarrow(s,t)$ is
$J=\partial_sx\partial_ty-\partial_tx\partial_sy=\cos\theta$.
Using these expressions we find the area element,
\begin{equation}
  \rmd A(s,t) = \Gamma^{1/2} \rmd s\rmd t =
  (1+q^2a^2\cos^2qt\sin^2\theta)^{1/2}\rmd s\rmd t,
\label{dA}
\end{equation}
and its projection onto the $xy$ plane,
\begin{equation}
  \rmd A_{\rm p}(s,t) = J\rmd s\rmd t = \cos\theta\rmd s\rmd t.
\label{dAp}
\end{equation}
An element of the boundary curve is given by $\rmd\vecR=(qa\cos
qt,1,0)\rmd t$, and the resulting arc-length element is
\begin{equation}
  \rmd R(s,t) = [1+(qa\cos qt)^2]^{1/2}\rmd t.
\label{dR}
\end{equation}
What is left to calculate is the local surface curvature, $c(s,t)$.
The local normal to the surface is given by
\[
  \vecn = \Gamma^{-1/2}(\partial_s\vecr\times\partial_t\vecr)=
  \Gamma^{-1/2} (-\sin\theta,qa\cos qt\sin\theta,\cos\theta).
\]
The mean curvature can then be calculated either from the trace of the
curvature tensor,
\[
  c(s,t) = -\half\tr (\partial n_i/\partial r_j),
\]
or by momentarily resorting to the Monge representation,
\[
  c = \nabla\cdot[\Gamma^{-1/2}\nabla z(x,y)],
\]
where the $\nabla$ operator is defined in the $xy$ plane. The
result is
\begin{equation}
  c(s,t) = \Gamma^{-3/2} [(1+q^2a^2\cos^2 qt)\dot{\theta} +
  q^2a\sin qt\sin\theta].
\label{cst}
\end{equation}




\vspace{1cm}

\vspace{1cm}

\begin{references}

\bibitem{Gelbart}
{\it Micelles, Membranes, Microemulsions, and Monolayers}, edited
by W. M. Gelbart, A. Ben-Shaul and D. Roux (Springer-Verlag, New
York, 1989).

\bibitem{Safran}
S. A. Safran, {\it Statistical Thermodynamics of Surfaces,
Interfaces, and Membranes} (Addison-Wesley, New York, 1994).

\bibitem{Birdi}
K. S. Birdi, {\it Lipid and Biopolymer Monolayers at Liquid
Interfaces} (Plenum Press, New York, 1989). K. S. Birdi, {\it
Self-Assembly Monolayer Structures of Lipids and Macromolecules at
Interfaces} (Kluwer Academic/Plenum Publishers, New York, 1999).

\bibitem{lung}
B. Robertson and H. L. Halliday, {Biochim. Biophys. Acta} {\bf
1408}, 346 (1998).

\bibitem{Glen}
G. D. Guttman and D. Andelman, {J. Phys. II France} {\bf 3}, 1411
(1993).

\bibitem{MJP}
S. T. Milner, J.-F. Joanny and P. Pincus, {Europhys. Lett.} {\bf
9}, 495 (1989).

\bibitem{Jalmes}
A. Saint-Jalmes, F. Graner, F. Gallet and B. Houchmandzadeh,
{Europhys. Lett.} {\bf 28}, 565 (1994).

\bibitem{Granek}
J.-G. Hu and R. Granek, {J. Phys. II France} {\bf 6}, 999 (1996).

\bibitem{multilayer1}
H. E. Ries and H. Swift, {Langmuir} {\bf 3}, 853 (1987).

\bibitem{multilayer2}
E. Hatta, H. Hosoi, H. Akiyama, T. Ishii and K. Mukasa,
{Eur. Phys. J. B} {\bf 2}, 347 (1998).

\bibitem{ejection}
P. Tchoreloff, A. Gulik, B. Denizot, J. E. Proust and F. Puisieux,
{Chem. Phys. Lipids} {\bf 59}, 151 (1991).

\bibitem{modulated1}
D. Andelman, F. Brochard, P.-G. de Gennes and J.-F. Joanny,
{C.R. Acad. Sci. Paris Ser. C} {\bf 301}, 675 (1985). D. Andelman
D., Brochard F. and J.-F. Joanny, {J. Chem. Phys.} {\bf 86}, 3673
(1987). D. Andelman, F. Brochard and J.-F. Joanny, {Proc. Natl.
Acad. Sci. USA} {\bf 84}, 4717 (1987).

\bibitem{modulated2}
M. Seul and D. Andelman, {Science} {\bf 267}, 476 (1995).

\bibitem{Leibler}
S. Leibler and D. Andelman, {J. Phys. France} {\bf 48}, 2013
(1987).

\bibitem{Japan}
D. Andelman D., T. Kawakatsu and K. Kawasaki, {Europhys. Lett.}
{\bf 19}, 57 (1992). T. Kawakatsu, D. Andelman, K. Kawasaki and T.
Taniguchi, {J. Phys. II France} {\bf 3}, 971 (1993). T. Taniguchi,
K. Kawasaki, D. Andelman and T. Kawakatsu, {J. Phys. II France}
{\bf 4}, 1333 (1994).

\bibitem{Wang}
Z. G. Wang, {J. Chem. Phys.} {\bf 99}, 4191 (1993).

\bibitem{Fred93}
F. C. MacKintosh and S. A. Safran, {Phys. Rev. E.} {\bf 47}, 1180
(1993).

\bibitem{Komura}
H. Kodama and S. Komura, {J. Phys. II France} {\bf 3}, 1305
(1993).

\bibitem{Harden}
J. L. Harden and F. C. MacKintosh, {Europhys. Lett.} {\bf 28}, 495
(1994).

\bibitem{German}
F. J\"{u}licher and R. Lipowsky, {Phys. Rev. E} {\bf 53}, 2670
(1996). P. B. Sunil Kumar and M. Rao, {Phys. Rev. Lett.} {\bf 80},
2489 (1998). W. T. G\'{o}\'{z}d\'{z} and G. Gompper, {Phys. Rev.
E} {\bf 59}, 4305 (1999). P. B. Sunil Kumar, G. Gompper and R.
Lipowsky, {Phys. Rev. E} {\bf 60}, 4610 (1999).

\bibitem{KaYee}
M. M. Lipp, K. Y. C. Lee, J. A. Zasadzinski and A. J. Waring, {\it
Science} {\bf 273}, 1196 (1996). M. M. Lipp, K. Y. C. Lee, A.
Waring and J. A. Zasadzinski, {Biophys. J.} {\bf 72}, 2783 (1997).
M. M. Lipp, K. Y. C. Lee, D. Y. Takamoto, J. A. Zasadzinski and A.
J. Waring, {Phys. Rev. Lett.} {\bf 81}, 1650 (1998).

\bibitem{Canay}
C. Ege and K. Y. C. Lee, unpublished.

\bibitem{ourEPL}
H. Diamant, T. A. Witten, A. Gopal and K. Y. C. Lee,
{Europhys. Lett.} {\bf 52}, 171 (2000).

\bibitem{ft_diffK}
One can just as well consider also differing bending moduli of the
two domains \cite{ourEPL}. The resulting expressions are more
cumbersome with essentially the same physics. For brevity we
consider in the current paper a single, uniform bending modulus
$K$.

\bibitem{ft_presstens}
The exact relation between the applied lateral pressure and
surface tension in a nonflat monolayer is a subtle issue
\cite{Granek}. In this paper we bypass it by referring to the
tension $\gamma$ alone, bearing in mind that it must decrease upon
increasing pressure.

\bibitem{ft_fluct}
We do not consider fluctuations of the surface; its equilibrium
shape is thus given by the minimum of the elastic energy. This
``zero-temperature" assumption is well justified in the current
case, as will be demonstrated in Sec.~\ref{sec_discussion}.

\bibitem{ft_kink}
We ignore in this paper the possibility of a sharp change in slope,
a crease, at the contact line. Such a crease would alter some of
our quantitative conclusions. Though such creases are consistent
with the symmetry of the system, we know of no estimate of their
magnitude.

\bibitem{ft_finite}
All of the above results, except for the precise profile
(Eq.~(\ref{prof_sharp})), hold also for a finite sheet clamped at
its ends (\ie maintaining the boundary conditions of vanishing
$\theta$ and $\dot{\theta}$ at both ends).

\bibitem{ft_linetension}
Here the bare line tension $\tau$ is assumed to contain all the
energetic contributions proportional to the boundary
length, including both short-range and electrostatic interactions;
see also Ref.~\cite{McConnell}.

\bibitem{ft_nl2}
Note again that the relevance of this instability relies on the
nonlinearity of our calculation. In a linearized theory (assuming
$\theta_0\ll 1$) one would get for the reduction in line tension
$g\simeq (-\theta_0/2)K\dc$, which could not compete with the bare
line tension $\tau$.

\bibitem{LL}
L. D. Landau and E. M. Lifshitz, {\it Theory of Elasticity}
(Butterworth-Heinemann, Oxford, 1986), Chap. II.

\bibitem{CP}
J. J. Binney, N. J. Dorwick, A. J. Fisher and M. E. J. Newman,
{\it The Theory of Critical Phenomena} (Oxford University Press,
Oxford, 1993).

\bibitem{Fisher}
M. E. Fisher, {Phys. Rev.} {\bf 176}, 257 (1968).

\bibitem{Onsager}
L. Onsager, {Phys. Rev.} {\bf 65}, 117 (1944).

\bibitem{Widom}
B. Widom, {J. Chem. Phys.} {\bf 43}, 3892 (1965).

\bibitem{Israelachvili}
J. N. Israelachvili, {\it Intermolecular \& Surface Forces}
(Academic Press, London, 1991), Chap. 11.

\bibitem{Vogel}
W. R. Schief, L. Touryan, S. B. Hall and V. Vogel, {J. Phys. Chem
B} {\bf 104}, 7388 (2000).
W. R. Schief, S. B. Hall and V. Vogel,
{Phys. Rev. E} {\bf 62}, 6831 (2000).

\bibitem{visco}
R. Miller, R. W\"ustneck, J. Kr\"agel and G. Kretzschmar,
{Colloid Surf. A} {\bf 111}, 75 (1996).

\bibitem{Ajay}
A. Gopal and K. Y. C. Lee, unpublished.

\bibitem{Knobler}
K. J. Stine and C. M. Knobler, {Phys. Rev. Lett.} {\bf 65}, 1004
(1990).

\bibitem{ltension1}
P. Muller and F. Gallet, {Phys. Rev. Lett.} {\bf 67}, 1106 (1991).

\bibitem{ltension2}
D. J. Benvegnu and H. M. McConnell, {J. Phys. Chem.} {\bf 96},
6820 (1992).

\bibitem{ltension3}
S. Rivi\`ere, S. H\'enon, J. Meunier, G. Albrecht, M. M.
Boissonnade and A. Baszkin, {Phys. Rev. Lett.} {\bf 75}, 2506
(1995).

\bibitem{ltension4}
S. Wurlitzer, P. Steffen and T. M. Fischer, {J. Chem. Phys.} {\bf
112}, 5915 (2000).

\bibitem{McConnell}
D. J. Keller, J. P. Korb and H. M. McConnell, {J. Phys. Chem.}
{\bf 91}, 6417 (1987). H. M. McConnell and V. T. Moy, {J. Phys.
Chem.} {\bf 92}, 4520 (1988). H. M. McConnell, {J. Phys. Chem.}
{\bf 94}, 4728 (1990). H. M. McConnell, {J. Phys. Chem.} {\bf 96},
3167 (1992). K. Y. C. Lee and H. M. McConnell, {J. Phys. Chem.}
{\bf 97}, 9532 (1993).

\bibitem{Seul}
M. Seul and J. Sammon, {Phys. Rev. Lett.} {\bf 64}, 1903 (1990).
M. Seul, {J. Phys. Chem.} {\bf 97}, 2941 (1993).

\bibitem{Goldstein}
R. E. Goldstein and D. P. Jackson, {J. Phys. Chem.} {\bf 98}, 9626
(1994).

\bibitem{DiffGeom}
See, \eg Ref.~\cite{Safran}, Chap. 1.

\end{references}
\end{document}